\documentclass[prb,twocolumn,preprintnumbers,amsmath,amssymb]{revtex4-1}
\usepackage{graphicx}

\begin{document}
\title{   High Chern number topological superfluids and new class of topological  phase transitions of Rashba spin-orbit coupled fermions
 on a lattice  }
%\title{   New Topological superfluid and phase transitions of spin-orbit coupled fermions with an attractive interaction in a square lattice  }
\author{  Yu Yi-Xiang$^{1,2}$, Fadi, Sun $^{2,3,4}$, Jinwu Ye $^{2,3,4}$ and Ningfang Song $^{1}$ }
\affiliation{
  $^{1}$  School of Instrument Science and Opto-electronics Engineering, Institute of Optics and Electronics, BeiHang University, Beijing 100191, China  \\
   $^{2}$  Department of Physics and Astronomy, Mississippi State  University, P. O. Box 5167, Mississippi State, MS, 39762   \\
$^{3}$ Department of Physics, Capital Normal University,
Key Laboratory of Terahertz Optoelectronics, Ministry of Education, and Beijing Advanced innovation Center for Imaging Technology,
Beijing, 100048, China   \\
$^{4}$ Kavli Institute of Theoretical Physics, University of California, Santa Barbara, Santa Barbara, CA 93106  }
\date{\today }

% Near the $ U(1) $ limit, $ \beta $ is small.
% Near the $ Z_2 $ limit, $ 1-\beta $ is small.
% The combination of the two methods provides rather complete physical picture from the $ U(1) $  to $ Z_2 $ limits.

\begin{abstract}
    Searching for the first topological superfluid (TSF) remains a primary goal of modern science.
    Here we study the system of attractively interacting fermions hopping in a square lattice  with
    any linear combinations of Rashba or Dresselhaus spin-orbit coupling (SOC) in a normal Zeeman field.
%    This is one of the simplest and most promising experimental setups to search for typological superfluids
%    and associated Majorana fermions in cold atom systems.
%    Here, we focus on half filling case.
    By imposing self-consistence equations at half filling, we find there are 3 phases: Band insulator ( BI ), Superfluid (SF) and Topological superfluid (TSF)
    with a Chern number $ C=2 $. The $ C=2 $ TSF happens in small Zeeman fields  and  weak interactions
    which is in the experimentally most easily accessible regime.
    The transition from the BI to the SF is a first order one due to the multi-minima structure of the ground state energy landscape.
    There is a new class of topological phase transition from the SF to the $ C=2 $ TSF at the low critical field $ h_{c1} $,
    then another one from the $ C=2 $ TSF to the BI at the upper critical field $ h_{c2} $.
    We derive effective actions to describe the two new classes of topological phase transitions,
    then use them to study the Majorana edge modes and the zero modes inside the vortex core of the $ C=2 $ TSF
    near both $ h_{c1} $ and $ h_{c2} $, especially explore their spatial and spin structures.
    We find the edge modes decay into the bulk with oscillating behaviors and determine both the decay and oscillating lengths.
    We compute the bulk spectra and map out the Berry Curvature distribution in momentum space near both $ h_{c1} $ and $ h_{c2} $.
    We also elaborate some intriguing bulk-Berry curvature-edge-vortex correspondences.
    Experimental  implications in both 2d non-centrosymmetric  materials under a periodic substrate and cold atoms in an optical lattice are given.
%    such as heating issue, temperature requirements and  detections of these new topological phenomena are discussed.
\end{abstract}

\maketitle

\section{ Introduction }
Since the experimental discovery of topological insulators \cite{kane,zhang} and Weyl semi-metals
\cite{weylrev,weylexp1,weylexp2a,weylexp2b,weylexp2c,weylexp3}, it became a primary goal to
find a first topological superfluid (TSF) \cite{read,kane,zhang} in any experimental systems.
The system of attractively interacting Rashba spin-orbit coupled fermions in a Zeeman field \cite{zhang} was considered to be one of the most promising systems to experimentally realize a topological superfluid.
It  was theoretically studied in the context of the hetero-structure made of
s-wave superconductor- noncentro-symmetric semiconductor- magnetic insulator (SM-SC-MI) \cite{das1,das2,das3,das4}.
In this SC-SM-MI hetero-structure,  the noncentro-symmetric semiconductor (SM) hosts a strong Rashba or Dresselhaus SOC,
the superconductor provides the S-wave pairing to the SM due to its proximity effects, the MI induces a Zeeman field applied to the SM.
Under the combined effects of the SOC, S-wave paring and the Zeeman filed, the SM sandwiched between the SC and MI
may enter into a Chern number $ C=1 $ TSF phase which hosts Majorana fermions in its vortex core.
Unfortunately,  so far  the experimental results on the hetero-structure came out as negative.
However, it was well known that a lattice system may offer a new platform to host new phases and phase transitions.
In this work, we study the system of attractively interacting fermions at half filling
subject to the Rashba spin-orbit coupling (SOC) hopping in a square lattice in a normal Zeeman field Eq.\ref{ham}.
%It can be viewed as a lattice regularization  of such a SC-SM-MI hetero-structure.
This system was first investigated in \cite{japan} in the context of cold atoms loaded on an optical lattice
and found to be a promising system to search for TSF  in cold atom systems.
Unfortunately, the self-consistent equations were ignored in \cite{japan}, so
what are the ground states and phase transitions can not be determined, possible experimental implications are rather limited.
%In order to avoid unwanted orbital effects, an alternative setup with the SM in the proximity to an
%s-wave superconductor subject to an in-plane magnetic field was proposed in \cite{ali}.
In the cold atom systems \cite{blochrev},  it is difficult to construct such a SC-SM-MI hetero-structure, but one advantage over the structure
is the absence of any orbital effects due to the charge neutrality of the cold atoms.
Another advantage is that  all relevant parameters are experimentally tunable.
For example, the negative interaction $ -U $ can be induced by a S-wave Feshbach resonance \cite{blochrev}.
The Rashba SOC and the Zeeman field $ h $  can be generated by  Raman laser scheme or optical lattice clock scheme \cite{expk40,expk40zeeman,2dsocbec,clock,clock1,clock2,SDRb,ben}.
The number of atoms $ N $ can be easily controlled. A crucial question to ask is what are the experimental conditions to
observe a possible TSF in such a lattice system ? If so, what are the properties of the TSF and associated topological phase transitions  ?
To answer these questions, one must impose the self-consist equations under the tunable experimental parameters such as
the SOC strength and the parameters $ N, U, h $ to determine the ground states and phase transitions.
% which can be directly detected with various techniques established
%in the cold atom experiments \cite{radio,mitdiss,topoexp,newexp2,edgeimage,dosexp}.
We will try to achieve this goal  in this paper.

\begin{figure}[tbp]
\includegraphics[width=8.5cm]{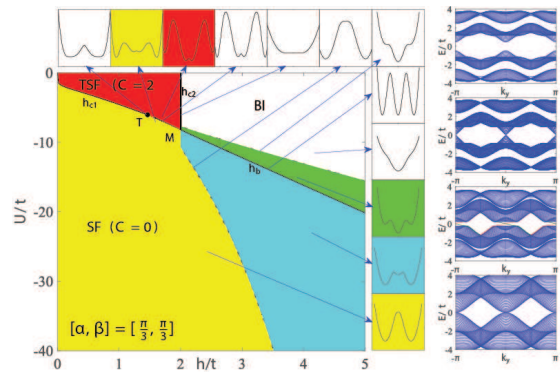}
%\hspace{0.2cm}
%\includegraphics[width=5.5cm]{combnsf.eps}
%\hspace{0.2cm}
%\includegraphics[width=3.85cm]{hcomp.eps}
%\includegraphics[width=5.0cm]{expsub.eps}
\caption{  The phase diagram of attractively interacting fermions with the Rashba SOC in a Zeeman $ h $ at half filling.
The SOC parameter is $[\protect\alpha ,\protect\beta ]=[\frac{\protect\pi }{3},\frac{\protect%
\pi }{3}]$.
The insets are the ground state energy $ E_{G} $  versus the bosonic SF order parameter $ \Delta $ at each phase and phase boundary.
%At the BI to the SF phase boundary $ h_b $, the three minima at $ 0 $ and $ \pm \Delta $ become degenerate, so it is a first order one.
The transition from the BI to the SF at $ h_b $ is a first order bosonic one with the superfluid order parameter $ \Delta $.
%where the four SF minima $ \pm \Delta_{+} $ and $ \pm \Delta_{-} $ become degenerate.
The SF to $ C=2 $ TSF transition at $ h_{c1} $ is a fermionic ( topological ) one with no order parameters.
Along $ h_{c1} $, it is third order from 0 to the topological Tri-critical point (T), then become first order from the T to
the multi-critical point ( M ).
The transition from the $ C=2 $ TSF to the BI at $ h_{c2} $  has both bosonic and fermionic ( topological ) nature.
The three lines $ h_{c1}, h_{c2} $ and $ h_{b} $ meet at the multi-critical point $ M $. The $ h_{c2} $ is strictly straight, while $ h_{c1} $
is tangent to $ h=0 $ axis near the origin.
On the right are the bulk and edge states in various phases.
(a) in the SF, there is a quasi-particle gap due to the pairing.
(b) Along $ h_{c1} $, there are two bulk gapless Dirac points at $ (0,\pi) $ and $ (\pi,0) $.
(c) the two edge states at $ k_y=0,\pi $ in the $ C=2 $ TSF.
(d) Along $ h_{c2} $, there are quadratic band touching at $ (0,0) $ and $ (\pi,\pi) $.
In the BI, there is a band gap due to the Zeeman field ( not shown ). }
%\hspace{0.2cm} }
%At large $ U $, the SF will crossover to BEC superfluid. }
\label{f33mini}
\end{figure}

In this paper, we find that it is very important to impose the self-consistency conditions which lead to the global phase diagram in Fig.\ref{f33mini}.
For the isotropic Rashba limit $ \alpha=\beta $, there are three phases: a topological superfluid phase (TSF) with a high Chern number $ C=2 $
at a small $ h $ and small $ U $, a Band insulator ( BI ) at a large $ h $ and
a normal SF at a large $ U $. The transition from the BI to the SF at $ h=h_b $ is a bosonic one with the pairing amplitude $ \Delta $
as the order parameter.
It is a first order one with meta-stable regimes on both sides of the transition ( denoted by the two dashed lines ) in Fig.\ref{f33mini}. The topological transition from the SF to the $ C=2 $ TSF at $ h=h_{c1} $ in Fig.\ref{f33mini} is a fermionic one at the two Dirac points
$ (\pi,0) $ and $ (0, \pi) $ with no order parameter.
It is a third order TPT in the first segment along $ h_{c1} $, then turn into a first order one at the
topological tri-critical point $ T $, continue to the multi-critical point $ M $.
The transition from the $ C=2 $ TSF to the BI at $ h_{c2} $ has both bosonic and fermionic nature,
the bosonic sector has the pairing amplitude $ \Delta $ as the order parameter representing the onset of the off-diagonal
long range order of the $ C=2 $ TSF,
the fermionic sector happens at the two Dirac points $ (0,0) $ and $ (\pi, \pi) $, representing the onset of the topological nature
of the $ C=2 $ TSF.
%The two sectors become critical at the same time at $ h=h_{c2} $.
The Berry curvature of the $ C=2 $ TSF in momentum space
are sharply peaked at $ (0,\pi) $ and $ (\pi,0) $ near $ h_{c1} $ in Fig.\ref{f33berryhc1}a,
but are located around $ (0,0) $ and $ (\pi,\pi) $ near $ h_{c2} $ with the non-trivial structure shown in Fig.\ref{f33berryhc2}a.
In the $ C=2 $ TSF, there are always $ C=2 $ Majorana edge modes at $ k_y=0 $ and $ k_y=\pi $ respectively which decay into the bulk
with an oscillating behavior.  The two Majorana edge modes carry both spins near $ h_{c1} $,
but spin up and spin down  at $ k_y=0 $ and $ k_y=\pi $ respectively near $ h_{c2} $.
There are  $ C=2 $  Majorana zero modes inside a vortex core which also show different spin structures near $ h_{c1} $ and $ h_{c2} $.
There are intriguing bulk energy spectrum-Berry curvature-edge state-vortex core relations.
We also discuss its  experimental  realizations in both 2d non-centrosymmetric  materials under a periodic substrate
and cold atoms in an optical lattice.
As a by-product, we also classify the possible 2d non-interacting topological effective actions.

 We consider the Hamiltonian of interacting two pseudo-spin (labeled as $\uparrow $ and $\downarrow $)
fermions hopping in the 2D square lattice subject to any linear combinations of Rashba and Dresselhaus SOC and a Zeeman field:
\begin{eqnarray}
H &= &-t\underset{\boldsymbol{i}}{\sum }\left[ c_{\boldsymbol{i}%
}^{\dagger }e^{i\alpha \sigma _{x}}c_{\boldsymbol{i+}\widehat{\boldsymbol{x}}%
}+c_{\boldsymbol{i}}^{\dagger }e^{i\beta \sigma _{y}}c_{\boldsymbol{i+}%
\widehat{\boldsymbol{y}}}+h.c.\right] -\mu \underset{\boldsymbol{i}}{\sum }c_{\boldsymbol{i}%
}^{\dagger }c_{\boldsymbol{i}} \notag \\
&- & h\underset{\boldsymbol{i}}{\sum }c_{\boldsymbol{i}}^{\dagger }\sigma_{z}c_{\boldsymbol{i}}
 + U \underset{\boldsymbol{i}}{\sum }c_{\boldsymbol{i}%
\uparrow }^{\dagger }c_{\boldsymbol{i}\downarrow }^{\dagger }c_{\boldsymbol{i%
}\downarrow }c_{\boldsymbol{i}\uparrow }
\label{ham}
\end{eqnarray}%
where $c_{\boldsymbol{i}}^{\dagger }=\left[
\begin{array}{cc}
c_{\boldsymbol{i}\uparrow }^{\dagger } & c_{\boldsymbol{i}\downarrow
}^{\dagger }%
\end{array}%
\right] $, $\sigma _{x,y,z}$ are three Pauli matrices, and $\widehat{%
\boldsymbol{x}}$ and $\widehat{\boldsymbol{y}}$ denote the unit vector in $x$
and $y$ direction respectively. The negative interaction $U<0$ can be tuned by
the Feshbach resonance in cold atoms or superconducting proximity effects in materials.
The chemical potential $\mu $ should be determined by the
filling factor $\nu =\frac{N}{2L_{x}L_{y}}$, where $N$ is the number of
fermions, $L_{x}$ (or $L_{y}$) is the size of the system along $x$ (or $y$)
direction, and the factor $2$ comes from the two spin species.

  The rest of the paper is organized as follows. In Sec.II, we present exact symmetry analysis on the Hamiltonian or its mean field form
  which will guide our qualitative physical pictures of the global phase diagram Fig.1.
  Then we explore the SF to the $ C=2 $ TSF transition at $ h_{c1} $ in Sec.III and the $ C=2 $ TSF to the BI transition at $ h_{c2} $ in Sec.IV.
  Then we use the effective actions near $ h_{c1} $ and $ h_{c2} $ derived in the previous two sections to study
  the spin and spatial structures of the edge modes in Sec.V and inside a vortex core in Sec.VI.
  In Sec. VII, we discuss the experimental realizations and detections of the $ C=2 $ TSF in
  both 2d non-centrosymmetric  materials under a periodic substrate and cold atoms in an optical lattice.
  In the final Sec.VIII, we reach conclusions and elaborate several perspectives.
  Some technical details are given in the 5 appendixes. Especially, in appendix E, we classify all the
  possible 2d non-interacting topological effective actions.

\section{ Exact symmetry analysis and Qualitative physical pictures. }

The Zeeman field breaks the Time reversal symmetry.
As usual, there is always a P-H symmetry on the BCS mean field Hamiltonian Eq.\ref{nambu}:
$ C H_{MF} ( \boldsymbol{k}) C^{-}= -H_{MF} ( -\boldsymbol{k}) $ which picks up
the four P-H invariant momenta $ (0,0),(\pi,0),(0,\pi), (\pi,\pi) $.
The Hamiltonian Eq.\ref{ham} also has the  $ {\cal P}_z $ symmetry:
$ k_x \rightarrow - k_x, S^{x} \rightarrow - S^{x}, k_y \rightarrow - k_y, S^{y} \rightarrow - S^{y}, S^{z} \rightarrow S^{z} $
which is also equivalent to a joint $ \pi $ rotation of the spin and orbital  around $ \hat{z} $ axis\cite{rh}.
This symmetry indicates $ \sigma_z H ( \boldsymbol{k}) \sigma_z= H ( -\boldsymbol{k}) $.
It also picks up the same four $ P_z $ symmetric  momenta \cite{pfaffian}.

At the isotropic Rashba limit $ \alpha=\beta $, it has the enlarged $ [ C_4 \times C_4 ]_D $ symmetry
which is also equivalent to a joint $ \pi/2 $ rotation of the spin and orbital  around $ \hat{z} $ axis.
This symmetry indicates the equivalence between $ (0,\pi) $ and $ (\pi,0) $.
The Hamiltonian is also invariant under $ \alpha \rightarrow \pi - \alpha, k_x \rightarrow  \pi - k_x  $ and
$ \beta \rightarrow \pi - \beta, k_y \rightarrow \pi - k_y  $.
At the extremely anisotropic limit $ (\alpha=\pi/2, \beta ) $, it indicates the  equivalence  between
$ (0,0) $ and $ (\pi,0) $, also between $ (0,\pi) $ and $ (\pi,\pi) $.

By introducing the superfluid order parameter $ \Delta $, we performed the mean field
calculations on the Hamiltonian Eq.\ref{ham}. By imposing the self consistent equations,
we determine $ \mu $ and $ \Delta $ in terms of given $ N $ and $ ( h, U ) $. The details are given in the method section.
In this manuscript, we only focus on the half-filling case $\mu =0$.
At the half filling $ \mu=0 $, the spectrum Eq.\ref{excitationE} has the symmetry $  E ( \boldsymbol{k})= E [  (\pi,\pi) + \boldsymbol{k} ] $
which indicates the energies at the four momenta split into two groups $ (0,0), (\pi,\pi) $ and $ (0,\pi),(\pi,0) $.
At the extremely anisotropic limit $ (\alpha=\pi/2, \beta ) $, the energies at the two groups become degenerate.
Any deviation from the anisotropic limit splits the 4 degenerate minima into the two groups
which opens a window for the TSF. The TSF window reaches maximum at the isotropic Rashba  limit $ \alpha=\beta  $
where the symmetry is  enlarged to $ [C_4 \times C_4]_D $. The main results for $ \mu=0 $ are shown in Fig.\ref{f33mini}.
%In the following, we describe  Fig.\ref{f33mini} in the clockwise version:
%from the BI to the SF at $ h_b $, then to the $ C=2 $ TSF at $ h_{c1} $,
%then back to the BI at $ h_{c2} $.

%It was shown in \cite{rh}, there is a enlarged spin-orbital coupled $ U(1)_{soc} $ symmetry along the line $ (\alpha=\pi/2, \beta) $
%which dictates the degeneracy at the 4 minima $ (0,0),(\pi,\pi),(0,\pi),(\pi,0) $.
%However, any deviation from the line spoils the $ U(1)_{soc} $ symmetry, so the the 4 degenerate minima splits into two groups
% $ (0,0),(\pi,\pi) $ and  $ (0,\pi),(\pi,0) $ dictated by the three  spin-orbital coupled discrete symmetries: $ {\cal P_x}, {\cal P_y}, {\cal P_z} $,
% the splitting opens a window for the TSF. The regime reaches maximum at the isotropic Rashba or    $ \alpha=\beta  $ where the symmetry is  enlarged to
% $ [C_4 \times C_4]_D $.

One can understand some qualitative features near $ h_{c2} $ in  Fig.1 starting from the non-interacting
SOC fermion spectrum ( namely the helicity basis ) with no pairing $ \Delta=0 $.
At $ h=0 $ axis in Fig.1, due to FS nesting, any $ U<0 $ leads to a trivial SF.
At $  h_{c2}= 2t( \cos \alpha + \cos \beta ) $, there is a quadratic band touching at $ (0,0) $ and $ (\pi,\pi) $
where there is already a gap  $ h-2t(  \cos \alpha - \cos \beta )=4t \cos \beta $ opening at $ (0,\pi) $ and $ (\pi, 0 ) $.
At $ 0 < h< h_{c2}= 2t( \cos \alpha + \cos \beta ) $, due to  the finite density of state (DOS ) at the FS $ \mu=0 $,
a weak $ U < 0 $ leads to a TSF with the $ p_x + i p_y $ pairing across 2 FS with the same helicity leading to $ C=2 $.
Here one gets a $ C=2 $ TSF almost for free: at a small $ h $ and a small attractive interaction $ U < 0 $.
At $ h=h_{c2} $, the FS disappears, there is only quadratic band touching at $ (0,0) $ and $ (\pi,\pi) $ with a zero DOS,
so one need a finite $ U_c $ to drive to a trivial SF. This explains why the $ h=h_{c2} $ is a straight line ending at $ U_c $ at the M point
in Fig.1.
%Using the vanishing DOS at the quadratic touching, one should be able to evaluate $ U_c $ analytically.
%In your Fig.2, It seems one can see a multi-critical point between 3 phases:  TSF, SF and  band insulator.
%Setting $ h_{c1}=h_{c2} $ leading to the maximum gap of the TSF $ \Delta= 4t \sqrt{ \cos \alpha \cos \beta } $.
%then plugging in this gap to the gap equation should lead to another $ U_c $,
When $ h> h_{c2} $, there is a band gap due to the Zeeman field at $ \mu=0 $, so one need even a larger $ U_c $ to reach a SF.
It turns out to be a 1st order transition to a trivial SF at $ h_b $ due to a jumping of the superfluid order parameter $ \Delta $,
so  it is a bosonic transition with gapped  fermionic excitations on both sides of the transition.
The first order transition may lead to a possible " phase separation" between the SF and BI in the two metastable regimes shown in Fig.1.
Obviously, it is the SOC which leads to the multi-minima structure of the ground state energy landscape leading to the first order BI-SF transition.
In fact, as shown in \cite{rhtran}, the SOC also leads to multi-minima structure of magnon spectrum in spin-orbital correlated magnetic phases.
So it is a generic feature for the SOC to lead to multi-minima structures in both the ground state and the excitation spectra.

Indeed, at $ \mu=0$, $\xi^2(\boldsymbol{k}_{0}) $ in Eq.M16
split into two groups: $\xi ^{2}\left( 0,0\right) =\xi ^{2}\left( \pi ,\pi \right) =4t^{2}\left( \cos \alpha +\cos
\beta \right) ^{2} > \xi ^{2}\left( 0,\pi \right) =\xi ^{2}\left(\pi, 0
\right) =4t^{2}\left( \cos \alpha -\cos \beta \right) ^{2}$.
If neither $ \alpha $ nor $\beta $ equals to $\frac{\pi }{2}$, the two groups take two different values
which divide the system into 3 different phases: SF, TSF and BI phase.
At $ h < h_{c1}=\sqrt{\xi^{2}\left( 0,\pi \right) +\Delta ^{2}} $,
it is in a trivial SF phase where the fermionic excitation energy $E_{\boldsymbol{k}-}$ is gapped in the bulk with no edge states.
There is a SF to TSF transition at $ h_{c1} $ where
the $E_{\boldsymbol{k}-}$ touches zero linearly and simultaneously at the two Dirac points at $\boldsymbol{k}=\left( \pi ,0\right) $ and $ \left(
0,\pi \right) $  shown in Fig.1a.
Then a second topological transition from the TSF to BI at $ h_{c2}= |\xi\left( 0,0\right)| =|\xi \left( \pi ,\pi \right) |
=2t\left( \cos \alpha +\cos\beta \right) $ where $ \Delta=0 $ and
the $E_{\boldsymbol{k}-}$ touches zero quadratically and simultaneously at $\boldsymbol{k}=\left( 0 ,0\right) $ and $ \left(
0,\pi \right) $  in Fig.1c. Inside the TSF $ h_{c1} < h < h_{c2} $, the bulk is gapped with two pairs of gapless edge
states at $ k_y=0, \pi $ on the boundaries of a finite-size system in Fig.1b.
In the BI $ h > h_{c2} $, it has a bulk gap due to the Zeeman field in Fig.1d.

If either $\alpha $ or $\beta $ is $\frac{\pi }{2}$ ( assuming $\alpha =\frac{\pi }{2}$ ), the two groups take the same value
$ \xi^2( \boldsymbol{k}_{0} )=4t^{2}\cos ^{2}\beta =h_{c}^{2} $ which divides
the system into only 2 different phases as shown in Fig.\ref{f26}. The TSF phase is squeezed to zero.
At $ h < h_{c} $, it is in the trivial SF phase with a bulk gap and no edge states.
At $ h= h_c $, the excitation energy $E_{\boldsymbol{k}-}$ touches zero quadratically and simultaneously at all four points  shown
in the inset of Fig.\ref{f26}. At $ h > h_{c} $, it gets into the gapped BI phase.

The main text focus on the isotropic Rashba limit $ \alpha=\beta $ where
 $ \xi ^{2}\left( 0,\pi \right) =\xi ^{2}\left(\pi, 0 \right) = 0 $.
In the anisotropic limit $ \alpha \neq \beta $, $ h_{c1} $ increases, the TSF phase shrinks ( Fig.\ref{f36} ).
In the extremely anisotropic limit $ \alpha=\pi/2 $, $ h_{c1}=h_{c2}=h_c $, the TSF phase shrinks to zero and disappears ( Fig.\ref{f26} ).
They will be discussed in details in the appendix B and C respectively.

\section{ The SF to the TSF transition at $ h=h_{c1} $.  }
In the bulk, the transition is driven by the gap closing of the two Dirac fermions  at  $ (0,\pi) $ and $ ( \pi, 0 ) $  with the same chirality.
%How to distinguish the TSF with $ C=2 $ from the trivial one  with $ C=0 $ ?
Now  we derive the effective $ 2 \times 2 $ Hamiltonian $ H (\vec{k} ) $  to describe the TPT near  $ h=h_{c1} $.
 At the $ h_{c1}=\sqrt{\xi^{2}\left( 0,\pi \right) +\Delta ^{2}} =\Delta $  and at the  two  Dirac points $ (0,\pi) $ and $ (\pi, 0 ) $,
 following \cite{tqpt}, one can find a $ 4 \times 4 $ unitary matrix  $ S_{(0,\pi)} $ to diagonize the $ 4 \times 4 $ Hamiltonian Eq.\ref{nambu}:
 $ S^{\dagger}_{(0,\pi)} H_{(0,\pi)} S_{(0,\pi)}= (2h_{c1},-2h_{c1},0,0) $.
 Now one can expand the Hamiltonian around $ h_{c1} $ and
 also near the Dirac point $ (0,\pi) $ by writing $ \delta h=h-h_{c1} $ and $ \vec{k}= (0,\pi) + \vec{q} $,
 then separate the $ 4 \times 4 $ Hamiltonian $ \tilde{H}= S^{\dagger}_{(0,\pi)} H S_{(0,\pi)} $
 into $ 2 \times 2 $ blocks  $ \tilde{H}= \left[
\begin{array}{cc}
H_{H} & H_{C} \\
H_{C}^{\dagger } & H_{L}%
\end{array}%
\right] $
 with the fermion field $ \Phi(k)= S^{\dagger}_{(0,\pi)} \Psi(k)=(\phi_H, \phi_L) $.
 Projecting to the 2 component low energy spinor: $ \phi_{ L \boldsymbol{k}}  =\frac{1}{\sqrt{2}}\left [
\begin{array}{c}
c_{- \boldsymbol{k} \uparrow }^{\dagger } -c_{ \boldsymbol{k} \downarrow } \\
c_{- \boldsymbol{k} \downarrow }^{\dagger } -c_{ \boldsymbol{k}  \uparrow }%
%\label{phiLk}
\end{array}%
\right ] $ space, we find the effective Hamiltonian $ H_{(0,\pi)} = H_{L}-H_{C}^{\dagger }H_{H}^{-1}H_{C} $:
\begin{eqnarray}
 H_{(0,\pi)} & = & \left( \delta h-\frac{ t^{2}\cos ^{2}\alpha \left( q^2_x - q^2_y \right)^{2}}{ 2 \Delta }\right)\sigma_{3}
                            \nonumber  \\
 & + & 2t\sin \alpha  ( q_{x}\sigma _{1}+ q_{y}\sigma _{2} )
 \label{f33hc1}
\end{eqnarray}
 where the Dirac fermion mass $ M=\delta h= h-h_{c1} $ changes the sign across the TPT boundary $ h=h_{c1} $.
 Note that because the SF order parameter $ \Delta $ remains a constant across the TPT,
 so it is just a pure fermionic TPT inside the SF with the dynamic exponent $ z=1 $.

 Eq.\ref{f33hc1} can be cast into the form:
\begin{equation}
 H(\vec{q})= \epsilon( \vec{q}) + d_a ( \vec{q} ) \sigma_a, ~~d_a ( \vec{q} )= ( Aq_x, Aq_y, M(\vec{q}) )
 \label{dirac}
\end{equation}
 where $ \epsilon( \vec{q})=0 $ dictated by the P-H symmetry, $ A= 2t \sin \alpha $ and $  M(\vec{q} )= \delta h -B( q^2_x- q^2_y)^2 $ where $
  B= \frac{ t^{2}\cos ^{2}\alpha }{ 2\Delta} >0  $.
 The first Chern number is given by:
\begin{equation}
 C_1=\frac{1}{4 \pi} \int d q_x d q_y \hat{\bf{d}}
 \cdot ( \frac{\partial \hat{\bf{d}}}{ \partial q_x} \times \frac{\partial \hat{\bf{d}}}{ \partial q_y} )
 \label{winding}
\end{equation}
  where $ \hat{\bf{d}}(\vec{q}) = {\bf d}(\vec{q})/|{\bf d}(\vec{q})| $ is a unit vector and the integral is over
  the 2d BZ in the original lattice model, but the whole
  2d $ (k_x,k_y) $ plane in the continuum limit.
  In the TSF, $ \delta h/B > 0 $, $  C_1=1 $.
%  The two Dirac fermions lead to $ C=2 $.
  In the trivial SF, $ \delta h /B < 0 $ ,$ C_1=0 $.
  The mass $ M=\delta h $ changes sign at the TQCP and is the only relevant term.
  The $ - B( q^2_x-q^2_y)^2 $ term is dangerous leading irrelevant
  near the TQCP in the sense that it is irrelevant at the TQCP, but it is important on the two sides of the TQCP
  and decide the thermal Hall conductivity of the two phases \cite{hightc,hightc01,hightc02}.

  Following the procedures in \cite{tqpt}, we find
  the ground state energy shows a singularity at its third order derivative  at the transition, so it is a 3rd order TPT.

% It is important to stress that if changing $ - B( q^2_x-q^2_y)^2 $ to $ - B( q^2_x-q^2_y) $, then $ C_1=1 $ would change to $ C_1=1/2 $.
% Incorporating the other Dirac fermion at $ (\pi,0) $, it describes a TPT from $ C=-1 $ TSF to $ C=1 $  TSF
% with the same jump of the Chern number $ \Delta C= 2 $ when away from half filling \cite{un}.

%It will be useful to examine the bulk-edge correspondence near $ h_{c1} $ to be shown in Appendix E.

%Following \cite{tqpt}, using the effective Hamiltonian Eq.\ref{f33hc1}, one can calculate the ground state critical exponent and compare with the
%numerical calculations using the original 4-bands Hamiltonian Eq.\ref{nambu} in the whole BZ.

One can get the effective Hamiltonian at  $ (\pi,0) $ by changing $ \sigma_1 \rightarrow - \sigma_1, \sigma_2 \rightarrow - \sigma_2 $
in Eq.\ref{f33hc1} or equivalently $ A \rightarrow- A $ in Eq.\ref{dirac}, but still with the same $ \phi_L $, so the pairing
remains the $ p_x + i p_y $ form \cite{reverse}.
Then Eq.\ref{winding} shows $ C_{(\pi,0)}= C_{(0,\pi)} =1 $, so the total Chern number $ C=C_{(\pi,0)}+ C_{(0,\pi)}=2 $.
Of course, $ H_{(0,\pi)} $ and $ H_{(\pi,0)} $ are related by the $ [C_4 \times C_4]_D $ symmetry at $ \alpha=\beta $.
In fact, the topological Chern number of a given band is the integral of the Berry curvature in the
whole BZ shown in Eq.\ref{winding}.  Here we show that
the global topology  can be evaluated just near a few isolated P-H symmetric points in an effective Hamiltonian in the continuum limit.
Note that the topological $ Z_2 $ indices ( Pfaffian )  are also evaluated at some isolated symmetric points \cite{pfaffian}.
They determine the topology of the bands in the whole BZ.

%Of course, the effective Hamiltonian in near $ (0,\pi) $ or $ (\pi,0) $, so contain no information on $ (0,0) $ and $ (\pi,\pi) $.
%They can only be seen in the original $ 4 \times 4 $ Hamiltonian, they are not useful anyway.

%How to see the interference effects of the two Majorana fermion edge states,
%at $ k_y=0 $ and $ k_y = \pi $ in real space ? any even-odd effects ? density wave odd in the edges ?

\begin{figure}[tbp]
\includegraphics[width=4.15cm]{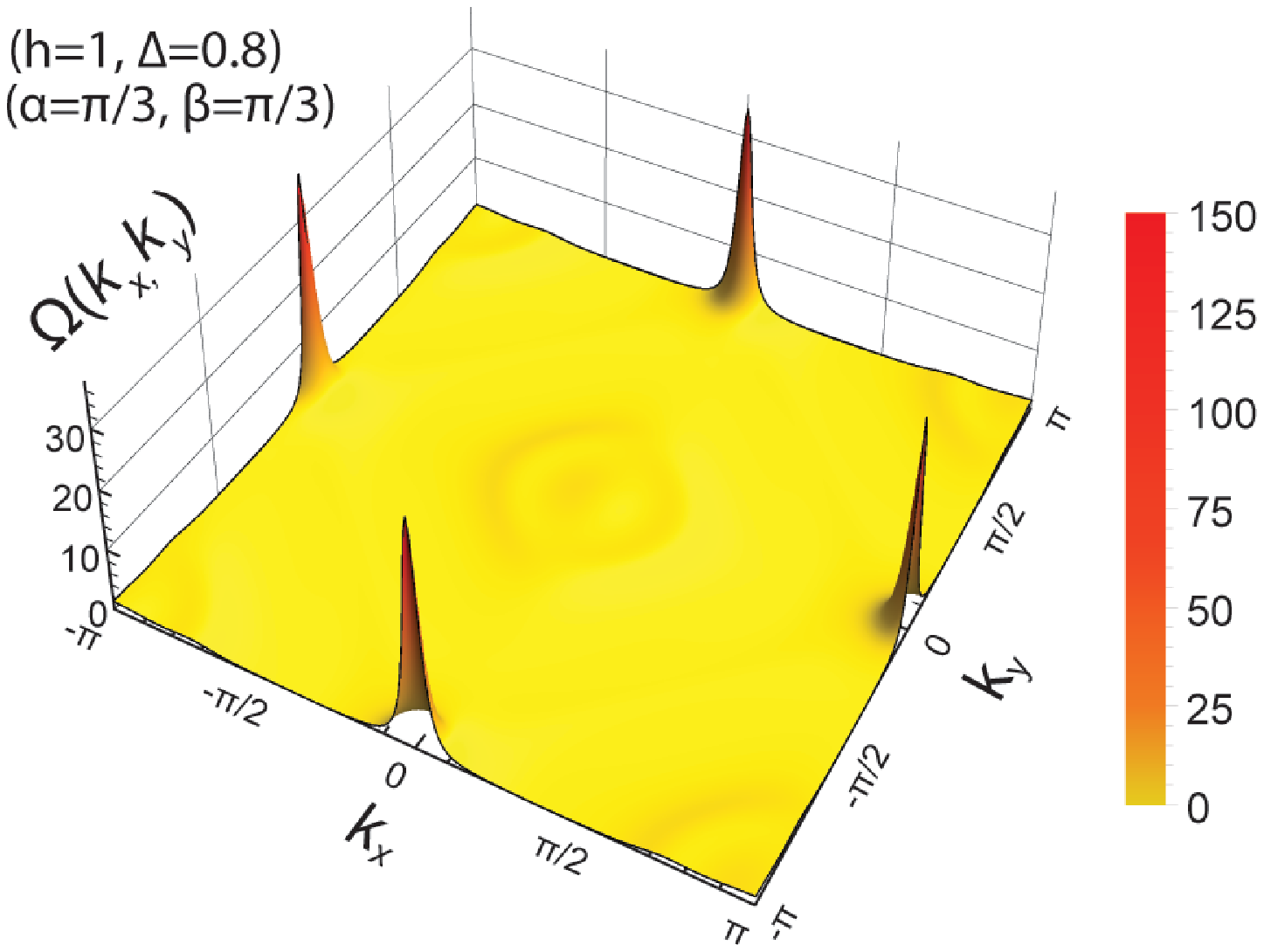}
\hspace{0.10cm}
\includegraphics[width=4.15cm]{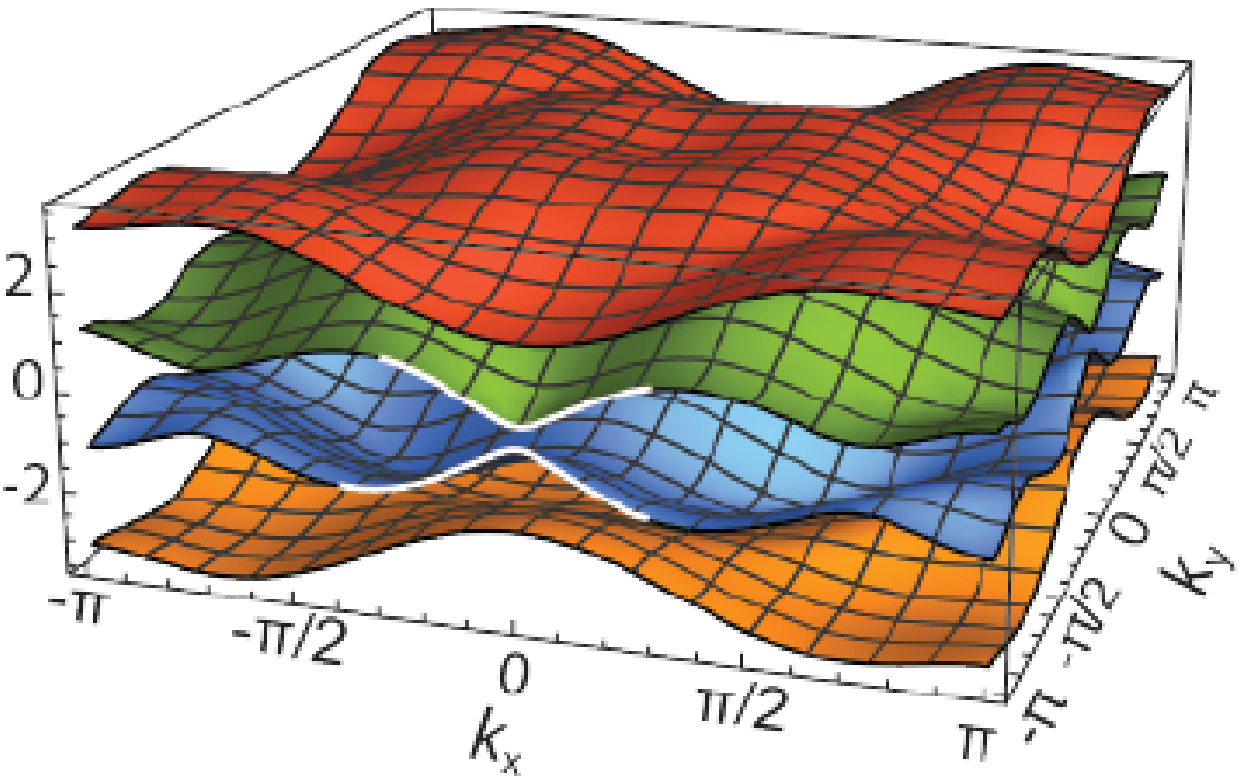}
\caption{(Color online) The Berry curvatures and energy bands  at $[\protect\alpha,
\protect\beta]=[\frac{\protect\pi}{3}, \frac{\protect\pi}{3}]$  near $ h_{c1} $ with $ ( h=1, \Delta=0.8) $ inside the TSF.
(a) The Berry curvature is sharply peaked at $ (0,\pi) $ and $ (\pi,0) $.
(b) The four quasi-particle energy bands. The two middle bands have a minimum gap denoted by the wide lines
    which leads to the Berry curvature in (a) near $ (0,\pi) $ and $ (\pi,0) $. }
\label{f33berryhc1}
\end{figure}

Using the original 4 bands theory and the three different methods outlined in the appendix D,
we calculated the Berry Curvature of $ E_{-}( \boldsymbol{q} ) $
in the whole BZ in Fig.\ref{f33berryhc1}a and find that near $ h_{c1} $,
they are sharply peaked at $ (0,\pi) $ and $ ( \pi,0) $. The corresponding 4 energy bands  are also shown in Fig.\ref{f33berryhc1}a.
These facts near  $ (0,\pi) $ and $ ( \pi,0) $ can be precisely captured by the 2 bands effective theory Eq.\ref{f33hc1}.
For example, from Eq.\ref{f33hc1}, one can determine the energy of the two middle bands:
\begin{equation}
 E_{\pm}( \boldsymbol{q} )=\pm \sqrt{\delta h^{2}+4t^{2}\sin ^{2}\alpha \left( q_{x}^{2}+q_{y}^{2}\right) }
\label{gaphc1}
\end{equation}
which has a minimum at $ \boldsymbol{q}=(0,0) $ as shown in Fig.\ref{f33berryhc1}b.
It leads to the Berry curvature distribution near  $ (0,\pi) $ and $ ( \pi,0) $  shown in Fig.\ref{f33berryhc1}a.

\section{  The TSF to the BI transition at $ h=h_{c2} $ }
Near $ h_{c2}=2t ( \cos \alpha + \cos \beta) $,  there are quadratic band touching at  $ (0,0) $ and $ ( \pi,\pi) $ as shown in Fig.\ref{f33mini}c.
Following similar procedures as those to derive the $ 2 \times 2 $ effective Hamiltonian Eq.\ref{f33hc1} near $ h_{c1} $ and $ (0,\pi) $, we derive
the $ 2\times 2 $ effective Hamiltonian near $ h_{c2} $ and $ (0,0) $:
\begin{eqnarray}
 H_{(0,0)} & = & \left( \delta h-t\frac{1+\cos ^{2}\alpha }{2\cos \alpha }\left(
q_{x}^{2}+q_{y}^{2}\right) \right )\sigma_{3}
                            \nonumber  \\
 & - & \frac{\Delta \tan \alpha }{2}\left( q_{x}\sigma _{1}+q_{y}\sigma _{2}\right)
\label{f33hc2}
\end{eqnarray}
where the two component low energy spinor $ \phi_{ L \boldsymbol{k}}  =\frac{1}{\sqrt{2}}\left [
\begin{array}{c}
c_{- \boldsymbol{k} \downarrow }^{\dagger }  \\
c_{ \boldsymbol{k} \downarrow }
%\label{phiLk}
\end{array}%
\right ] $ which contains only spin down. The Dirac fermion mass $ M=\delta h= h_{c2}-h $ changes its sign across the TPT boundary $ h=h_{c2} $.

 Eq.\ref{f33hc2} can also be cast into the form Eq.\ref{dirac}
 where $ \epsilon( \vec{k})=0 $, $ A= \frac{\Delta \tan \alpha }{2} $  and $  M(\vec{k} )= \delta h -B( q^2_x + q^2_y),
 B=t\frac{1+\cos ^{2}\alpha }{2\cos \alpha } >0 $.
 The first Chern number is still given by Eq.\ref{winding}:
 If $ M/B < 0 $ and $ \Delta=0 $ in the BI, $ C_1=0 $.  It becomes a gapped non-relativistic fermion due to the Zeeman field:
 $  E(k) \sim [ h-h_{c2}] + B ( q^2_x+ q^2_y) $.
 If $ M/B > 0 $ and $\Delta \neq 0 $ in the TSF, $  C_1=1 $.  There is also a gap opening due to the effective  $ p_x +i p_y $  pairing
 $ \Delta $. It has the dynamic exponent $ z=2 $ at the QCP $ \delta h=0, \Delta=0 $.
 So there are two relevant operators: the mass term $ M= \delta h $  and the $ p_x + i p_y $ pairing term $ \Delta $.
 The $ - B( q^2_x + q^2_y) $ term is dangerous leading irrelevant near the QCP in the sense that it is irrelevant at the QCP,
 but it is important to the physical properties of the two phases on the two sides of the QCP.

%However, at the QCP, $ A=0, M=0 $, but $ B >0 $, so the spectrum becomes quadratic $  E(k)=  B ( q^2_x+ q^2_y) $.

\begin{figure}[tbp]
\includegraphics[width=4.15cm]{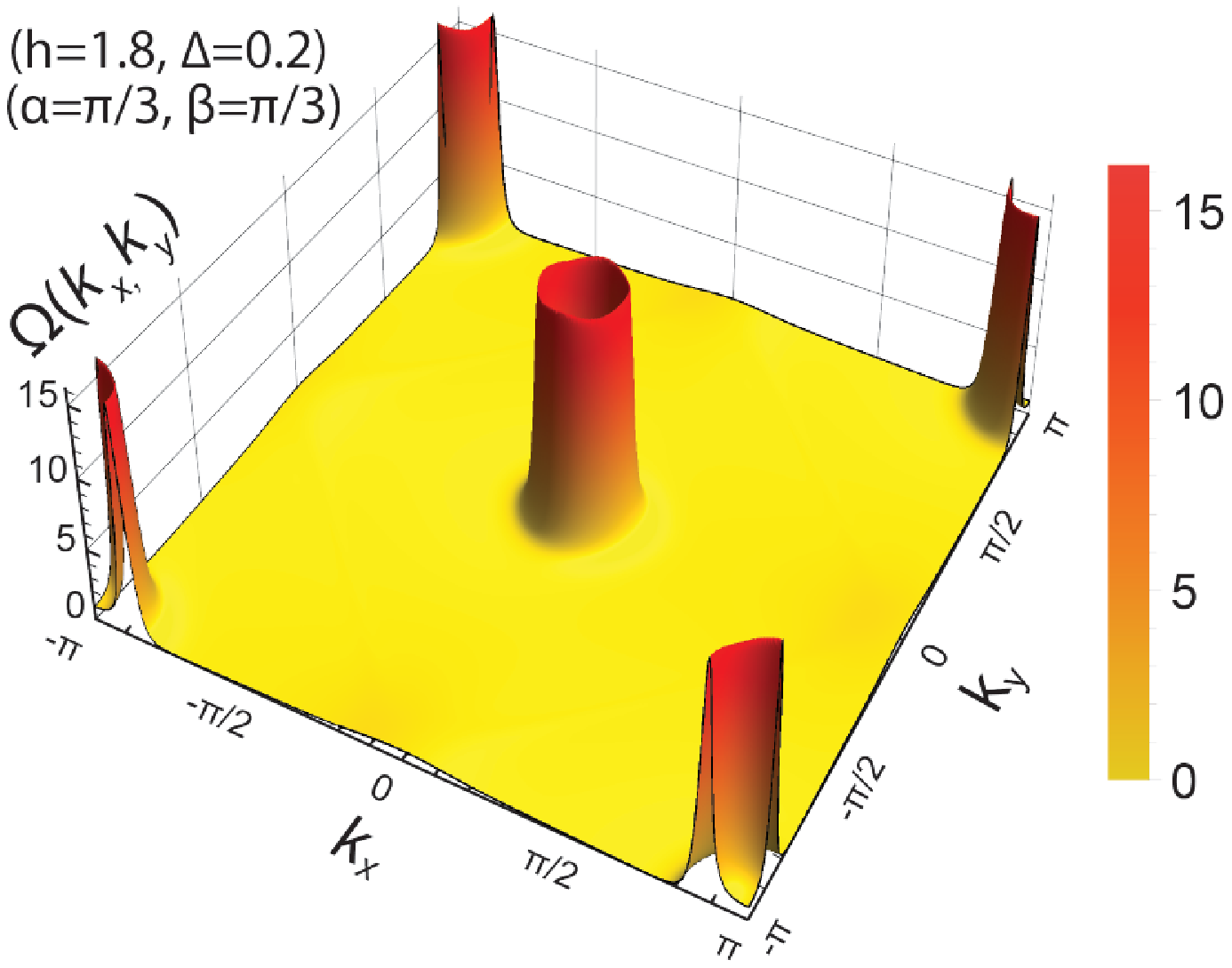}
\hspace{0.10cm}
\includegraphics[width=4.15cm]{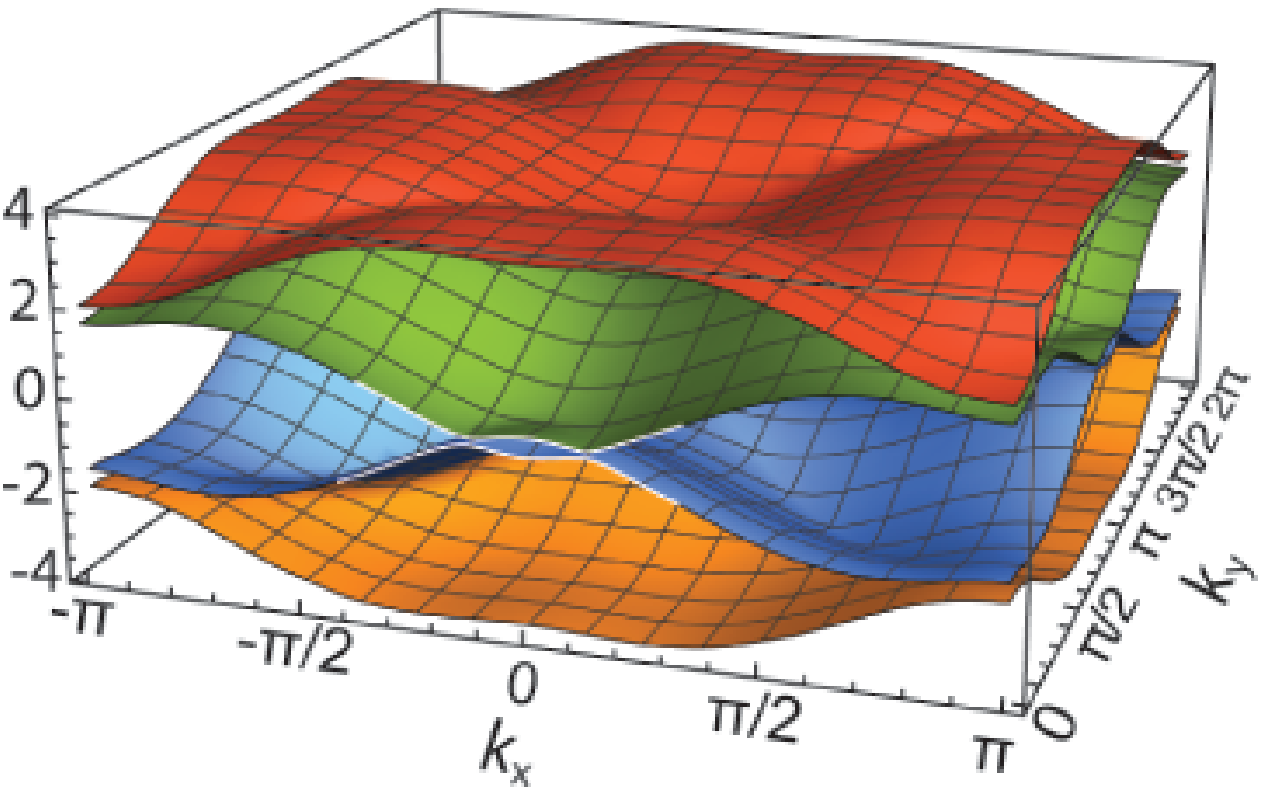}
\caption{(Color online) The Berry curvatures and energy bands  at $[\protect\alpha,
\protect\beta]=[\frac{\protect\pi}{3}, \frac{\protect\pi}{3}]$ near $ h_{c2} $ with $ ( h=1.8, \Delta=0.2) $ inside the TSF.
(a) The Berry curvature has a dip at $ (0,0) $ and $ (\pi,\pi) $, but peaked at a ring around the two Dirac points.
(b) The four quasi-particle energy bands. The two middle bands have the shape denoted by the wide lines near $ (0,0) $ and $ (\pi,\pi) $.
  The energy gap contour of the two middle bands  are nearly circular which leads to
  the Berry curvature structure in (a). This bulk feature leads to the oscillating behavior of the edge state wavefunction when decaying into the bulk
  shown in the method section. }
\label{f33berryhc2}
\end{figure}

One can get the effective Hamiltonian at  $ (\pi,\pi) $ by changing $ \sigma_1 \rightarrow - \sigma_1, \sigma_2 \rightarrow - \sigma_2 $
in Eq.\ref{f33hc2} or equivalently $ A \rightarrow -A $ in Eq.\ref{dirac}. Especially, it has a different
low energy two component spinor $ \phi_{ L \boldsymbol{k}}  =\frac{1}{\sqrt{2}}\left [
\begin{array}{c}
c_{ \boldsymbol{k} \uparrow }   \\
c_{- \boldsymbol{k} \uparrow }^{\dagger }
%\label{phiLk}
\end{array}%
\right ] $ which contains only spin up.
Then Eq.\ref{winding} shows $ C_{(\pi,\pi)}= C_{(0,0)} =1 $,
so the total Chern number $ C=C_{(\pi,\pi)}+ C_{(0,0)}=2 $ in the TSF. So the distribution of the Berry curvature is moving
from $ (0,\pi) $ and $ (\pi,0) $ near $ h_{c1} $ as shown in Fig.\ref{f33berryhc1}a to $ (0,0) $ and $ (\pi,\pi) $ near $ h_{c2} $ shown in Fig.\ref{f33berryhc2}a.

Indeed, using the original 4 bands theory, using three different methods ( See appendix D),
we calculated the Berry Curvature of $ E_{-}( \boldsymbol{q} ) $
in the whole BZ in Fig.\ref{f33berryhc2}a and find
they are localized around $ (0,0) $ and $ ( \pi,\pi) $ near $ h_{c2} $ shown in Fig.\ref{f33berryhc2}a.
The corresponding 4 energy bands  are also shown in Fig.\ref{f33berryhc2}b.
These facts can be precisely captured by the 2 bands effective theory Eq.\ref{f33hc2}.
For example, one can determine the bulk energy of the two middle bands:
\begin{equation}
   E_{\pm}( \boldsymbol{q} ) = \pm  \sqrt { \left[ M- B \left( q_{x}^{2}+q_{y}^{2}\right) \right]^2 + A^2 \left( q_{x}^{2}+q_{y}^{2}\right) }
\label{gaphc2}
\end{equation}
    which means that in the TSF side $ M= \delta h > 0 $, if assuming $ C=2MB-A^2 >0 $, then it  has a maximum at $ \boldsymbol{q}=(0,0) $ and
    a minimum at $ q^2=C/2B^2 $ with a minimum gap $ E_{min}= \frac{A \sqrt{4 MB-A^2}}{2 B} >0 $.
    This is indeed the case as shown in Fig.\ref{f33berryhc2}b.
    It leads to the non-trivial Berry curvature distribution near  $ (0,0) $ and $ ( \pi,\pi) $  shown in Fig.\ref{f33berryhc2}a.
    This non-trivial structure of the bulk gap is also crucial to explore the bulk-edge correspondence near $ h_{c2} $  in the next section.

The main difference between the TPT at $ h=h_{c1} $ described by Eq.\ref{f33hc1} and that at $ h_{c2} $ described by Eq.\ref{f33hc2}
is that in the former, the SF order parameter $ \Delta $ is non-critical across the SF to TSF transition at $ h_{c1} $,
the effective $ p_x+ip_y $ pairing amplitude in Eq.\ref{f33hc1} is
given by the SOC strength $ t \cos \alpha $, so it is a pure fermionic transition.
However, in the latter, the SF order parameter $ \Delta $ is also critical across the
TSF to the BI transition at $ h_{c2} $, the effective $ p_x+ip_y $ pairing amplitude in Eq.\ref{f33hc1} is given
by the S-wave pairing $ \Delta  $ multiplied by an SOC related factor $ \tan \alpha $.
So it involves two natures instead of just a pure fermionic transition:
(1) Conventional bosonic nature due to the superfluid order parameter $ \Delta $.
(2) Topological fermionic nature due to the Dirac fermions in the TSF side.
It happens near the two Dirac points $ (0,\pi) $ and $ (\pi,0) $ near $ h_{c1} $ in the Fig.\ref{f33mini}.
However, it moves to $ (0,0) $ and $ (\pi,\pi) $  near $ h_{c2} $.

\section{ Edge modes in the $ C=2 $ TSF, oscillation of the edge states and bulk-edge correspondences. }
\begin{figure}[tbp]
\includegraphics[width=4.25cm]{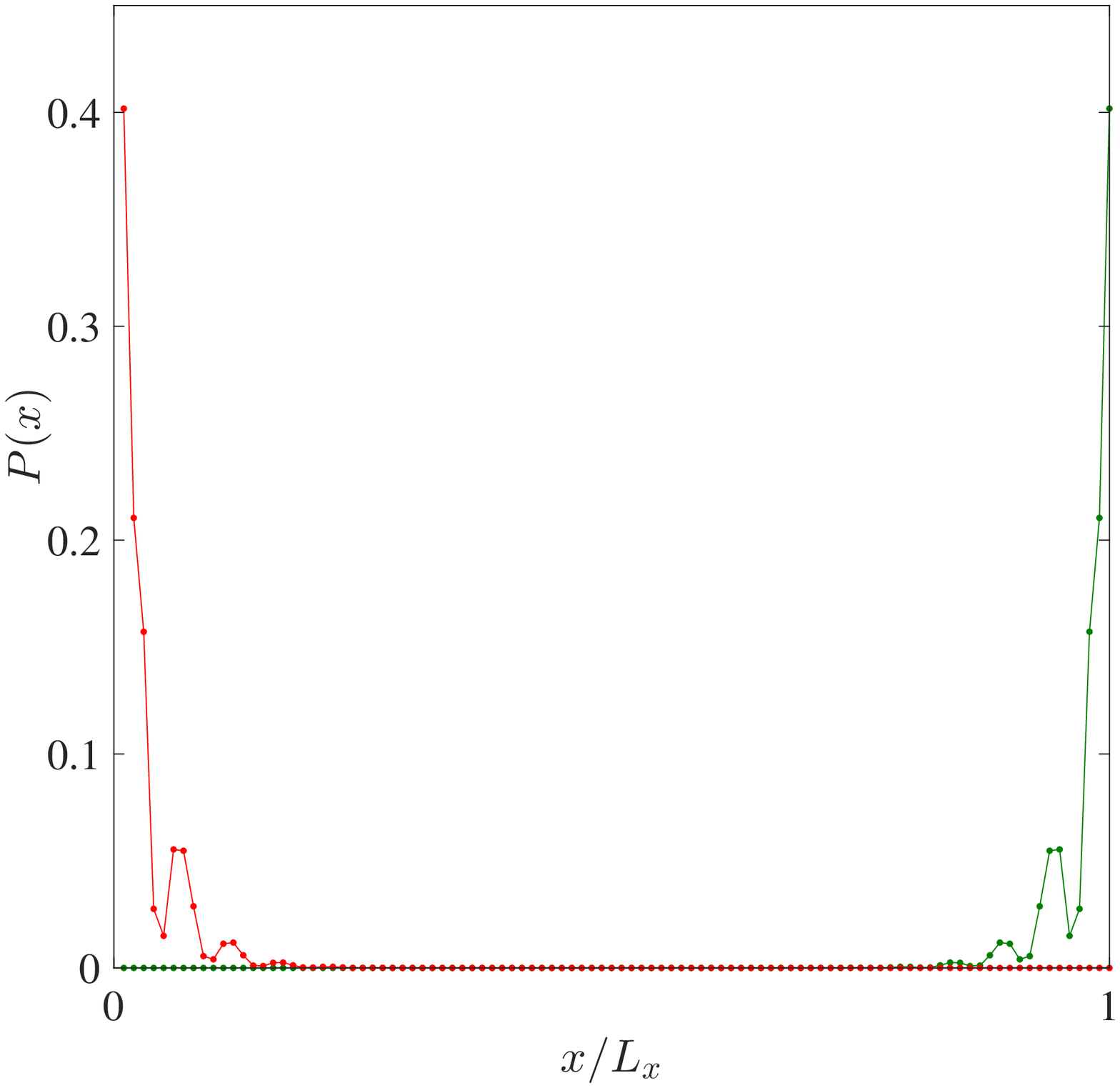}
\includegraphics[width=4.25cm]{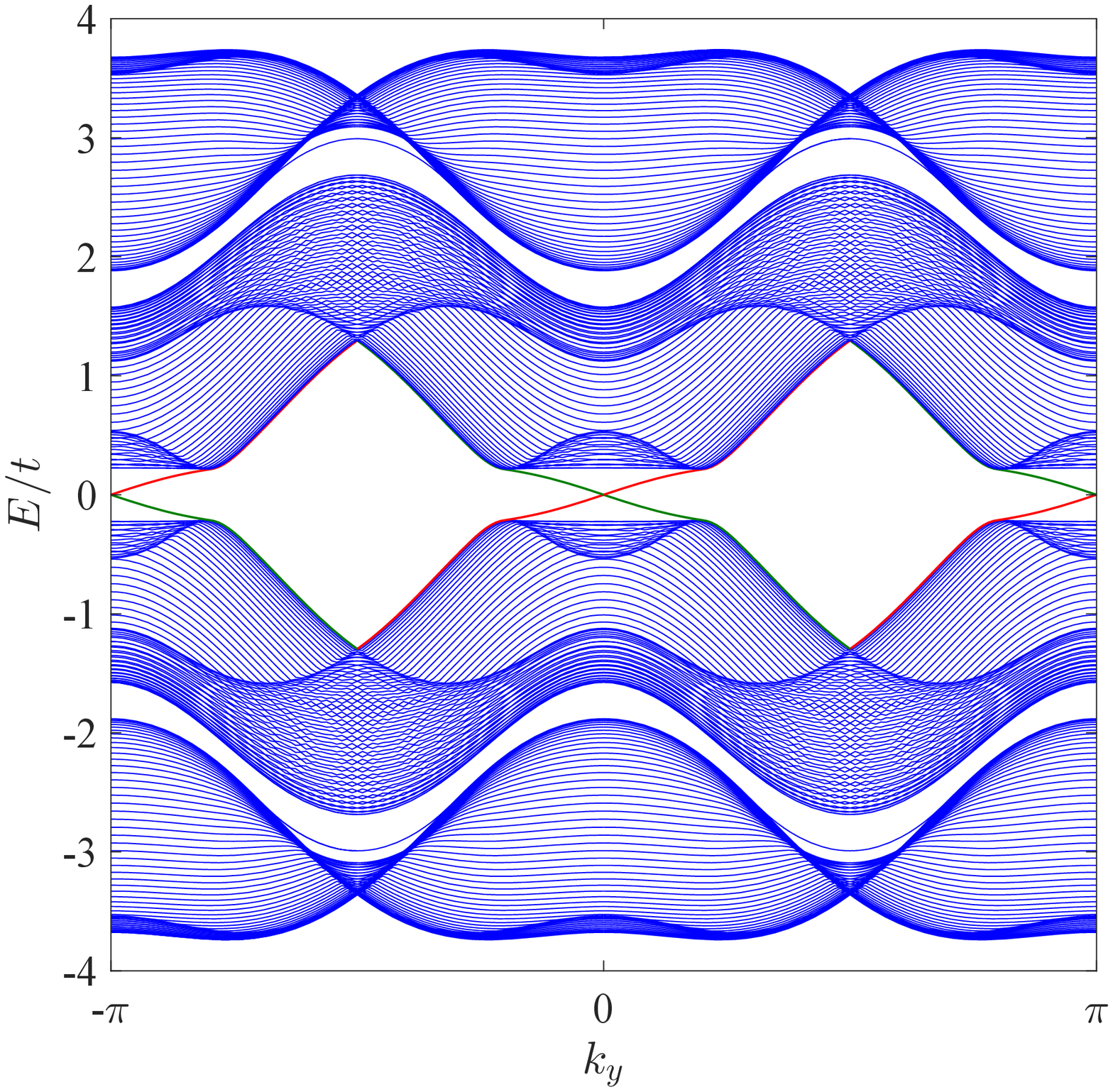}
\caption{(Color online) The normalized edge state wavefunction of the TSF  in both real and momentum space at
the two edges $ x=0, L $ and at $ k_y=0 $. The SOC parameters are $[\protect\alpha,
\protect\beta]=[\frac{\protect\pi}{3}, \frac{\protect\pi}{3}]$ and the $ (h, \Delta ) $ corresponding to
one point inside the TSF  near $ h_{c2} $  in Fig.\ref{f33mini}.
(a) In the real space, it shows oscillating behaviors towards decaying into the bulk.
The red and green lines show the two oppositely propagating edge modes on the two opposite sides of the sample.
(b) In the momentum space, it show two edge states at $ k_y=0, \pi $.  The red and green lines correspond to
the edge mode in the left and right in (a) respectively.  }
\label{f33wf}
\end{figure}

As shown in Fig.\ref{f33mini},\ref{f33wf}, using the open boundary conditions in the $ x $ directions,
the Exact Diagonization (ED) study near $ h_{c1} $ shows that in the TSF side with $ C=2 $,
there are two branches of Majorana fermion edge states  $ k_y=0 $ and $ k_y = \pi $.
In the trivial SF side with $ C=0 $, there is no edge states.
In the following, we analyze the two edge modes in the TSF from
the effective actions Eq.\ref{f33hc1} near $ h_{c1} $ and Eq.\ref{f33hc2}  near $ h_{c2} $ respectively.

{\sl 1. Spin and spatial structures of the Edge states near $ h_{c1} $. }

  As shown in Eq.\ref{f33hc1}, the effective $ p_x+ip_y $ pairing amplitude $ \Delta_e=2 t \sin \alpha $ is relatively large,
  so the TSF has a relatively large gap. The notable feature near $ h_{c1} $ is that Eq.\ref{f33hc1} is not isotropic in $ (q_x, q_y) $, so the edge
state depends on orientation of the edge.
Setting $ \tilde{q}_x= ( q_x-q_y )/\sqrt{2}, \tilde{q}_y= ( q_x+q_y )/\sqrt{2} $,
then making simultaneous  rotation in the spin space, Eq.\ref{f33hc1} can be rewritten as:
\begin{equation}
 \tilde{H}_{(0,\pi)}  = \left( \delta h-\frac{ t^{2}\cos ^{2}\alpha \tilde{q}^2_x  \tilde{q}^2_y }{ \Delta }\right)\sigma_{3}
 +  2t\sin \alpha  ( \tilde{q}_{x} \tilde{\sigma}_{1}+ \tilde{q}_{y} \tilde{\sigma}_{2} )
\label{f33hc1r}
\end{equation}
   So the edge state along the edge $ \tilde{x} =0 $ ( or $ x=\pm y $ )  can be similarly constructed
   in the rotated $ \tilde{} $ basis as in \cite{kane,zhang}.
   However, to see the detailed structure of the edge state wavefunctions
   $ u_{k_{y}}^{\ast }\left( x\right) =-v_{-k_{y}}\left( x\right) $
   such as decaying into the bulk with possible oscillations,
   one may need keep higher order terms in the bulk effective action Eq.\ref{f33hc1}.

   In the expression of $ \phi_{ L \boldsymbol{k}} $ listed above Eq.\ref{f33hc1}, $ k_y $ remain good quantum number,
   setting $ k_x \rightarrow x $ leads to
   the edge operator at a given $ k_y $:
\begin{equation}
c_{2L,k_{y}} ( x ) =\frac{1}{\sqrt{2}}\left [
\begin{array}{c}
c_{-k_{y}\uparrow }^{\dagger }( x ) -c_{k_{y}\downarrow } ( x ) \\
c_{-k_{y}\downarrow }^{\dagger }( x ) -c_{k_{y}\uparrow }( x )%
\label{phiLx1}
\end{array}%
\right ]
\end{equation}%
   We get the Majorana edge mode $ \gamma_2( k_{y} ) $ in Eq.\ref{edge}:
\begin{eqnarray}
\gamma_2(k_y) &=&\frac{1}{\sqrt{2}} \int^{\infty}_0 dx [ \left( u_{k_y}^{\ast}(x)
c_{-k_y\uparrow }^{\dagger }( x ) -v_{k_y}^{\ast}(x) c_{k_y \uparrow }( x ) \right)      \nonumber  \\
& + & \left( v_{k_y}^{\ast}( x ) c_{-k_y \downarrow }^{\dagger} ( x )
-u_{k_y}^{\ast }( x ) c_{k_y \downarrow }( x ) \right) ]     \nonumber   \\
 &=& \gamma_{2 \uparrow}(k_y) + \gamma_{2 \downarrow}(k_y)
\label{gamma2ky}
\end{eqnarray}%
   which satisfies $\gamma^{\dagger}_2(k_{y})=\gamma_2(-k_{y}) $ and includes both spin up $ \gamma_{2 \uparrow}(k_y) $ in the first line
   and the spin down $ \gamma_{2 \downarrow}(k_y) $ in the second line.

   Similarly, using the effective action $ H_{(\pi,0)} $, one can derive the  Majorana edge mode
   $ \gamma_1( k_{y} ) $ near $ k_y=0 $ in Eq.\ref{edge}.
   Because the unitary transformation $ S_{(\pi,0)}=S_{(0,\pi)} $, so the form of Eq.\ref{phiLx1} and Eq.\ref{gamma2ky}
   hold also for $ \gamma_1( k_{y} ) $.

%  As shown in the following, setting $ q_x \rightarrow -i \partial_x $ at any given $ q_y $ in the $ 2 \times 2 $ effective Hamiltonian Eq.\ref{f33hc1}
%  around $ (0,\pi) $ or $ (\pi,0 ) $, one can determine the Majorana edge mode $ \gamma_2 $ near $ k_y=\pi $ in Eq.\ref{gamma2ky}
%  and  $ \gamma_1 $ near $ k_y=0 $ respectively.
  In terms of the Majorana edge mode $ \gamma_2 $ near $ k_y=\pi $ in Eq.\ref{gamma2ky}  and  $ \gamma_1 $ near $ k_y=0 $,
  one can write the effective 1d Majorana fermion $ \gamma^{\dagger}_i=\gamma_i, i=1,2 $  edge Hamiltonian:
\begin{equation}
  H_{edge}= \int dy [-i v_f \gamma_i \partial_y \gamma_i ]
\label{edge}
\end{equation}
  where  $ i=1,2 $ stand for the two edge modes and $ v_f=\Delta_e=2 t \sin \alpha $ is the edge velocity near $ h_{c1} $.

  The original edge Majorana  fermion on the edge $ x=0 $ can be expressed in terms of the two edge modes:
\begin{equation}
  \psi_1(y)= \gamma_{1}(y) + (-1)^{y} \gamma_2(y)
\label{edgereal}
\end{equation}
  where $  \psi^{\dagger}_{1} (y)=\psi_1(y) $.
  So one can evaluate the  Majorana fermion correlation functions along the 1d $ x=0 $ edge Eq.\ref{edge} and \ref{edgereal}.

%   It is important to stress that despite there are two Majornana fermions edge modes in the momentum space, only
%  one Majorana edge mode $ \psi_1 $ in real space.
  In fact, one can define another Majorana edge mode $ \psi_2(y)= \gamma_{1}(y) + (-1)^{y+1} \gamma_2(y) $ which
  is decoupled, so plays no role.
  It is easy to check that for the two sets of Majorana fermions:
  $ \{ \gamma_i (y), \gamma_j(y^{\prime} ) \} =\delta_{ij} \delta(y-y^{\prime} ),
   \gamma^{\dagger}_i (y)=\gamma_i(y),  \psi^{\dagger}_i(y)=\psi_i(y),  \{ \psi_i(y), \psi_j(y^{\prime} ) \}
   =2 \delta_{ij}\delta(y-y^{\prime} ), i,j=1,2 $.
  The extra factor of $ 2 $ shows that the edge state contains 2 Majorana fermions \cite{norm}.
%  Note that Here, we are using the normalization  $ \{ \gamma_i, \gamma_j \} =\delta_{ij} $
%  instead of the more conventional one $ \{ \gamma_i, \gamma_j \} =2 \delta_{ij} $
%  which seems more natural in the BdG equation normalization.

{\sl 2. The Spin and spatial structures of Edge states near $ h_{c2} $. }

As shown in Eq.\ref{f33hc2}, the effective $ p_x+ip_y $ pairing amplitude $ \Delta_e= \Delta \tan \alpha/2 $ is relatively small,
so the TSF has a relatively small gap. To get the edge  modes along a edge at $ x=0 $  near $ h_{c2} $, we set
 $ q_x \rightarrow -i \partial_x $ in Eq.\ref{f33hc2}.
 Similar procedures as in \cite{zhang} can be used to find the edge mode at  a given $ k_y $ near $ h_{c2} $
 by imposing the additional Majorana fermion condition $ u_{k_{y}}^{\ast }\left( x\right) =-v_{-k_{y}}\left( x\right) $.
 The $ k_y=0 $  energy eigenvalue equation near $ h_{c1} $ is:
\begin{equation}
   M + B \lambda ^{2} +  A\lambda =0
\label{eigen2}
\end{equation}
 where as written below Eq.\ref{f33hc2}:
  $ A=\frac{\Delta \tan \alpha }{2}, B= t\frac{1+\cos ^{2}\alpha }{2\cos \alpha }, M=\delta h $.

 One salient feature here is that both $ A $ and $ M $ are critical near $ h_{c2} $.
 By a simple GL analysis, $ \Delta \sim ( \delta h )^{1/2} $, so, in general, $ D= A^{2}- 4 M B $
 could be either positive or negative. However, as shown in Fig.4 and the main text, it is negative here,
 so Eq.\ref{eigen2} has the two physical roos with negative real part:
\begin{equation}
   \lambda_{3,4}=-\frac{A \pm i  \sqrt{ |D| } }{2 B }
\end{equation}
  which we denote by  $ \lambda_3, \lambda_4= \lambda^{*}_3 $.

In a sharp contrast, near the TI to BI transition, $ M $ changes sign across the transition, so $ M $ is a small quantity,
while $ A $ remains un-critical across the transition, so $ D= A^2 -4 M B $ is always positive,
the edge state wavefunctions decay into the bulk monotonically  with no oscillations.

%there is no oscillation in the decay of the edge state into the bulk.

 After imposing the additional hard boundary condition $ \psi(x=0) =0 $, we can find a unique
 $ E=0 $ edge state wavefunction:
\begin{equation}
\left[
\begin{array}{c}
u_{0}\left( x\right)  \\
v_{0}\left( x\right)
\end{array}%
\right] = i R \left[
\begin{array}{c}
-e^{i\frac{\pi }{4}} \\
e^{i\frac{\pi }{4}}
\end{array}%
\right] \left( e^{\lambda _{3}x}-e^{\lambda _{3}^{\ast }x}\right)
\label{edgewf2}
\end{equation}%
  where  $ R=\sqrt{\frac{ |Re\lambda _{3}| | \lambda _{3} |^{2} }
 {2 ( |Re\lambda_{3} |^{2}+ | \lambda _{3} |^{2} ) } } $ is the normalization constant.

 The magnitude of the wavefunction $ |\Phi(x)|^2 $  Eq.\ref{edgewf2}
 decays into the bulk with the decaying length $ l^{-1}_d= |Re \lambda_3| =A/2B $ with oscillating
 period $ l^{-1}_o= | Im \lambda_3| = \sqrt{|D|}/2 B  $ which is consistent with the decaying-oscillating behaviors in the ED study in Fig.\ref{f33wf}. Again, the above procedures can be extended to derive the wavefunction at any given $ k_y  $
 with the eigen-energy $ E=v_f q_y $ where $ v_f= A $.
 When comparing with the bulk energy spectrum Eq.\ref{gaphc2},
 we find that it is the oscillating length  $ l^{-1}_o=E_{min}/v_f $ which can be expressed as the minimum bulk gap over the edge velocity,
 while the decay length $ l^{-1}_d=E_{min}/\sqrt{ |D| } $ is more complicated than that near $ h_{c1} $.
 Because $ \sqrt{ |D| }> A $, so $ l_o < l_d $. These analytical predictions are indeed observed in the ED results in Fig.\ref{f33wf}.
  Fig.\ref{f33wf}a shows the wavefunction of the edge mode at a given $ k_y $
  which was achieved by ED a $ 4 L_{x} \times 4 L_{x} $ matrix at any given $ k_y $.
  Notably, the edge state wavefunction at a given $ k_y $ decay into the bulk with some oscillating behaviors.

%  It is easy to see that the solution exists only in the TSF where $ \delta h > 0 $.

 In the expression of $ \phi_{ L \boldsymbol{k}} $ listed below Eq.\ref{f33hc2}, setting $ k_x \rightarrow x $ leads to
 the edge operator at a given $ k_y $:
\begin{equation}
c_{3L,k_{y}}\left( x\right) =\left[
\begin{array}{c}
c_{-k_{y}\downarrow }^{\dagger }\left( x\right)  \\
c_{k_{y}\downarrow }\left( x\right)
\end{array}%
\right]
\end{equation}%
   which contains only spin down. The Majorana edge mode is given by:
\begin{eqnarray}
\gamma_3( k_{y} )=\int dx\left[ u_{k_{y}}^{\ast }\left(
x\right) c_{-k_{y}\downarrow }^{\dagger }\left( x\right) +v_{k_{y}}^{\ast
}\left( x\right) c_{k_{y}\downarrow }\left( x\right) \right]
\label{gamma3ky}
\end{eqnarray}%
   which satisfies $\gamma^{\dagger}_3( k_{y} )=\gamma_3( -k_{y} ) $
   and includes the only spin down.

   Similarly, using the effective action $ H_{(\pi,\pi)} $, one find the Majorana fermion near $ k_y=\pi $:
\begin{equation}
c_{4L,k_{y}}\left( x\right) =\left[
\begin{array}{c}
c_{k_{y}\uparrow }\left( x\right)  \\
c_{-k_{y}\uparrow }^{\dagger }\left( x\right)
\end{array}%
\right]
\end{equation}%
   which contains only spin up and
\begin{eqnarray}
\gamma_4( k_{y} )=\int dx\left[ u_{k_{y}}^{\ast }\left(
x\right) c_{k_{y}\uparrow }\left( x\right) +v_{k_{y}}^{\ast
}\left( x\right) c_{-k_{y}\uparrow }^{\dagger }\left( x\right)
 \right]
\label{gamma4ky}
\end{eqnarray}%
   which satisfies $\gamma^{\dagger}_4( k_{y} )=\gamma_4( -k_{y} ) $
   and includes the only spin up.

%In the following, we used the effective action Eq.\ref{f33hc2} to derive the edge modes in the continuum limit.
%It was first studied in \cite{read} by ignoring the quadratic $  -B(q^{2}_x + q^{2}_y) $ term.
%When taking this quadratic term into account, it was computed in Ref.\cite{zhang} in the context of Topological Insulator (TI).
%The main difference is that Ref.\cite{zhang} is for TI, so the edge is the 1d helical edge state,
%Here it is a Majorana fermion edge state dictated by the PH symmetry.
%  Because the transition from the BI to the $ C=2 $ TSF involves both the onset
%  of the bosonic SF order $ \Delta $ and the fermionic topological order with the sign change of the Dirac fermion mass $ M $,
%  the edge eigenvalue have both real and imaginary part, the real part leads to
%  the decay length $ l_d $  into the bulk, while the imaginary part leads to
%  the oscillating length $ l_o < l_d $.

%  In sharp contrast, near the TI to BI transition in Ref.\cite{zhang}, there is only
%  the fermionic topological order with the sign change of the Dirac fermion mass $ M $,
%  the edge state wavefucntions decay into the bulk monotonically  with no oscillations.

In terms of the edge mode $ \gamma_3 $ near $ k_y=0 $ in Eq.\ref{gamma3ky}
and $ \gamma_4 $ near $ k_y=\pi $ in Eq.\ref{gamma4ky}, we also reach the same Eq.\ref{edgereal}
with $ v_f=\Delta_e= \Delta \tan \alpha/2,  i=3,4 $ and $ \psi_3(y)= \gamma_{3}(y) + (-1)^{y} \gamma_4(y) $.
The crucial differences than the two Majorana fermions $ \gamma_1, \gamma_2 $  near $ h_{c1} $ is that
$ \gamma_3 $ and $ \gamma_4 $ near $ h_{c2} $ contain only spin down and spin up respectively.

\section{ Majorana bound states inside a vortex core of the $ C=2 $  TSF }

 It was known that at a $ C=1 $ TSF, a $ n=\pm 1 $ vortex  holds one Majornan fermion zero mode \cite{read}.
 Here, we have a $ C=2 $ TSF in Fig.1 with two chiral edge modes,
% it remains interesting to study how a $ n=1 $ vortex will hold two Majornan fermion zero modes.
 In general, it is expected that the Chern number $ C=2 $ is equal to the number of edge modes
 and also the number of Majornan zero modes inside a $ n=\pm 1 $ vortex core.
 Similar to the study of the edge states, one can use the effective action near $ h_{c1} $ and $ h_{c2} $ to
 study  analytically the zero modes inside a S-wave vortex core.
 When introducing a vortex in the phase winding of the SF order parameter,
 it will affect most the low energy fermionic responses near $ (0, \pi) $ and $ (\pi,0) $ when $ h $ is near $ h_{c1} $
 or near $ (0, 0) $ and $ (\pi,\pi) $ when $ h $ is near $ h_{c2} $ respectively.
 The existence and stability of the zero modes are protected by the Chern number $ Z $ class classification of the TSF, so
 are independent of the continuum approximation made in the effective actions.
 Similar continuum approximations were used to study the quasi-particles in the vortex states
 of high $ T_c $ superconductors \cite{hightc,hightc01,hightc02}.

{\sl 1. Two Majorana zero modes $ \gamma_1, \gamma_2 $  near $ h_{c1} $:  }

In the $ 2 \times 2 $ effective Hamiltonian Eq.\ref{f33hc1} near $ (0,\pi) $ with $ C=1 $,  setting
$ \Delta \rightarrow \Delta_0 e^{i \theta} $, the first term remains intact, but the SOC strength in the second term
$ t \sin \alpha \rightarrow t \sin \alpha e^{i \theta} $  acquires an effective phase from the order parameter phase winding.
Setting $ q_x \rightarrow -i \partial_x, q_y \rightarrow -i \partial_y $ and paying special attentions to the anisotropy in the
$ -B(q^2_x-q^2_y)^2 $ term,
one may derive the wavefunctions $ ( u( \boldsymbol{r} ), v(\boldsymbol{r} ) ) $ satisfying $ u^{\ast}( \boldsymbol{r} )= -v(\boldsymbol{r} ) $.

 In the expression of $ \phi_{ L \boldsymbol{k}} $ listed above Eq.M4, setting $ \boldsymbol{k} \rightarrow  \boldsymbol{r} $ leads to
 the particle operator at a given $ \boldsymbol{r} $:
\begin{equation}
c_{2L} ( \boldsymbol{r} ) =\frac{1}{\sqrt{2}}\left [
\begin{array}{c}
c_{\uparrow }^{\dagger }( \boldsymbol{r} ) -c_{ \downarrow } ( \boldsymbol{r} ) \\
c_{\downarrow }^{\dagger }( \boldsymbol{r} ) -c_{ \uparrow }( \boldsymbol{r} )%
\label{phiLr1}
\end{array}%
\right ]
\end{equation}%
  which contain both spin up and spin down.
  It can be fused with the zero-mode wavefunctions $ ( u(\boldsymbol{r} ), v(\boldsymbol{r} )  ) $
  to lead to the Majorana zero mode $ \gamma_2 $ in Eq.\ref{corehc1}:
\begin{eqnarray}
\gamma_2 &=&\frac{1}{\sqrt{2}} \int d\boldsymbol{r} [ \left( u^{\ast}(\boldsymbol{r}  )
c_{\uparrow }^{\dagger }( \boldsymbol{r}  ) -v^{\ast}(\boldsymbol{r}  ) c_{ \uparrow }( \boldsymbol{r}    ) \right)      \nonumber  \\
& + & \left( v^{\ast}( \boldsymbol{r}  ) c_{ \downarrow }^{\dagger} ( \boldsymbol{r}  )
-u^{\ast }(\boldsymbol{r}) c_{ \downarrow }( \boldsymbol{r}    ) \right) ]   \nonumber  \\
&=& \gamma_{2 \uparrow} + \gamma_{2 \downarrow}
\label{gamma2r}
\end{eqnarray}%
   which satisfies $\gamma^{\dagger}_2=\gamma_2,  \gamma^2_2=1/2$ and includes both spin up
   $ \gamma_{2 \uparrow},  \gamma^2_{2 \uparrow}=1/4 $ in the first line
   and the spin down $ \gamma_{2 \downarrow},  \gamma^2_{2 \downarrow}=1/4 $ in the second line.

One can do a similar calculation near $ (\pi, 0 ) $ to get the second trapped Majorana zero mode $ \gamma_1 $ which also contains both spin up and spin down.
One may combine the two trapped Majorana zero modes inside a S-wave vortex core near $ h_{c1} $
into a single Dirac fermion:
\begin{equation}
 \psi_1= \gamma_1 + i \gamma_2
\label{corehc1}
\end{equation}
Its number $ \psi^{\dagger}_1 \psi_1=0,1 $ counts the occupations on the zero mode. Its exchange statistics is just a fermionic one.
There is no long-range entanglement between two distant vortices. A local operation can change the Dirac fermion occupation number
inside the vortex core.

%  In fact, one can define another Majorana zero mode $ \psi_2(= \gamma_{1} - \gamma_2 $ which
%  is decoupled, so plays no role.
%  It is easy to check that for the two sets of Majorana fermions \cite{norm}:
%  $ \{ \gamma_i, \gamma_j \} =\delta_{ij}, \gamma^{\dagger}_i=\gamma_i,  \Psi^{\dagger}_i=\Psi_i,  \{ \Psi_i, \Psi_j \}=2 \delta_{ij},  i,j=1,2 $.
%  The extra factor of $ 2 $ shows that the vortex holds 2 Majorana fermions.

{\sl 2. Two Majorana zero modes $ \gamma_3, \gamma_4 $ near $ h_{c2} $.  }

Similarly, in the $ 2 \times 2 $ effective Hamiltonian Eq.\ref{f33hc2} near $ (0,0) $ with $ C=1 $,  setting
$ \Delta \rightarrow \Delta_0 e^{i \theta} $, the first term remains intact, the second term
$ \Delta \tan \alpha \rightarrow \Delta_0 \tan \alpha e^{i \theta} $
acquires the phase and is nothing but a $ p_x+ i p_y $ pairing vortex.
The Majorana fermion zero mode inside such a cylindrical symmetric vortex core in the polar coordinate $ ( r, \theta) $
has been worked out in many previous literatures  \cite{read,japan,das1,das2,das3}.
Combining the known wavefucntions $ ( u( \boldsymbol{r} ), v(\boldsymbol{r} ) ) $ satisfying $ u^{\ast}( \boldsymbol{r} )= -v(\boldsymbol{r} ) $
with
\begin{equation}
c_{3L}\left( \boldsymbol{r} \right) =\left[
\begin{array}{c}
c_{ \downarrow }^{\dagger }\left(  \boldsymbol{r} \right)  \\
c_{ \downarrow }\left( \boldsymbol{r}   \right)
\end{array}%
\right]
\end{equation}%
   leads to
\begin{eqnarray}
\gamma_3=\int d\boldsymbol{r}  \left[ u^{\ast }\left(\boldsymbol{r}
     \right) c_{ \downarrow }^{\dagger }\left( \boldsymbol{r}   \right) +v^{\ast}\left(  \boldsymbol{r}  \right) c_{\downarrow }
     \left(  \boldsymbol{r}  \right) \right]
\label{gamma3r}
\end{eqnarray}%
   which satisfies $\gamma^{\dagger}_3=\gamma_3, \gamma^2_3=1/2 $ and includes the only spin down.

   Similarly, using the effective action $ H_{(\pi,\pi)} $, one can derive another Majorana zero mode $\gamma_4 $.
Combining the known wavefucntions with
\begin{equation}
c_{4L}\left( \boldsymbol{r} \right) =\left[
\begin{array}{c}
 c_{ \uparrow }\left( \boldsymbol{r}  \right) \\
 c_{ \uparrow }^{\dagger }\left(  \boldsymbol{r} \right)
\end{array}%
\right]
\end{equation}%
  lead to
\begin{eqnarray}
\gamma_4=\int d\boldsymbol{r}  \left[ u^{\ast }\left(\boldsymbol{r}
     \right) c_{\uparrow } \left(  \boldsymbol{r}  \right)
      +v^{\ast}\left(  \boldsymbol{r}  \right) c_{ \uparrow }^{\dagger }\left( \boldsymbol{r}   \right) \right]
\label{gamma4r}
\end{eqnarray}%
   which satisfies $\gamma^{\dagger}_4=\gamma_4, \gamma^2_4=1/2 $ and includes the only spin up.

Combining the two trapped Majorana zero modes Eq.\ref{gamma3r} and Eq.\ref{gamma4r} inside a S-wave vortex core near $ h_{c2} $
leads to a single Dirac fermion $ \psi_2= \gamma_3 + i \gamma_4 $.

%Unfortunately, some people say only when Chern number is odd, one gets Majornana fermions bound state in a vortex core.
%When  Chern number is even, no Majornana fermions bound state.  This need to be checked.
%But we do get two Majornana edge modes at $ k_y=0, \pi $.

It is instructive to compare Eq.M9 with Eq.\ref{corehc1} which leads to the following interesting edge-vortex core correspondence.
In the former, the two Majorana edge modes are separated by
the conserved momentum $ k_y= \pi $ along the $ x=0 $ edge, so their linear combination leads to the Majorana fermion $ \psi_1 $
with a twice magnitude.
While, in the latter, the two Majorana edge modes are trapped inside the same vortex core, so can be combined into
one Dirac fermion.

 There were previous studies on nearly zero modes inside a vortex core of a superconducting state in graphene \cite{nearzero}.
 There are four of them. However, these four zero modes  appear only in linear approximation,
 but are not protected by any topological indices, therefore can be lifted by lattice effects,
 in sharp contrast to the $ C=2 $ Majorana zero modes here which are protected by the $ Z $ class of TSF.

%Their wavefunctions may also different spatial dependencies as shown in Eq.\ref{edgewf1} and Eq.\ref{edgewf2}.

%When moving from $ h_{c1} $ to $ h_{c2} $ inside the TSF, we expect the edge mode will crossover from the equal spin superposition in $ \gamma_1 $
%to just spin down in $ \gamma_3 $ near $ k_y=0 $, the equal spin superposition in $ \gamma_2 $
%to just spin down in $ \gamma_4 $ near $ k_y=\pi $. Of course, the two edge modes stay near  $ k_y=0 $ and $ k_y=\pi $ as dictated by
%the symmetries of the Hamiltonian.

%  It corresponds to the edge Majorana fermion $ \Psi_1 $ in Eq.\ref{edgereal} instead of the
%  $ \psi_1 $ and $ \psi_2$ near $ k_y=0 $ and $ k_y=\pi $ respectively in Fig.\ref{f33wf}b.
%  Following Refs.\cite{read,japan,das123}, one may need to solve the BdG equation
%  with just one edge. Indeed, they see some complex roots which may correspond to the oscillations in Fig.5.

\section{ Experimental realization and detections of the TSF. }
  In condensed matter systems, as said in the introduction, any of the linear superpositions
  of the Rashba SOC $ k_x \sigma_x + k_y \sigma_y $ and Dresselhaus SOC $ k_x \sigma_x - k_y \sigma_y $ always exists
  in various noncentrosymmetric 2d or layered materials. In momentum space, such a linear combination
  $ \alpha k_x \sigma_x + \beta k_y \sigma_y $ can be written as the kinetic term
  in Eq.\ref{ham} in a periodic substrate. The anisotropy in the SOC parameter $ (\alpha,\beta ) $
  can be adjusted by the strains, the shape of the surface or gate electric fields.
  The negative strength $ U < 0 $ in Eq.\ref{ham} can be induced by the superconducting proximity effects.
  More simply, Eq.\ref{ham} can be viewed as the lattice regularization of the  SC-SM-MI hetero-structure.
  So all the phenomena in Fig.\ref{f33mini} can be observed in these 2d non-centrosymmetric materials.

%   It was known that the phases and phase transitions in a lattice are different and  much richer than those in a continuum.
%   However, so far, most of previous works studied the continuum of Eq.\ref{ham},
%   there are very few theoretical works on a lattice \cite{japan}.
%   The goal of this work is to predict the regime in $ ( h,U) $ and $ N $  to search for topological superfluids and the topological transitions to
%   its neighboring quantum phases, study the properties of the TSF, the nature of these transitions and their experimental detections.
   In cold atom systems, the chemical potential $ \mu $ is not measurable or controllable, only the number of atoms $ N $
   is, so the self-consistence equations must be imposed to get the realistic phases and phase transitions
   in $ ( h, U ) $ at a given $ N $  and to have any experimental impacts.
   In the present system, the BI only happens in a lattice. The TSF to BI transition  at $ h_{c2} $ in Fig.\ref{f33mini} only happens in a lattice.
   The TSF happens in the small $ h $ and small $ U $ in the Fig.\ref{f33mini} which
   is the experimentally most easily accessible regimes.
   This fact is very  crucial for all the current experiments \cite{expk40,expk40zeeman,2dsocbec,clock,clock1,ben}
   to probe possible  many body effects of SOC fermion or spinor boson gases.
%   because so far, the heating effects are controllable only at weak couplings, which get more problematic as the interaction gets stronger.

   The topological order of the TSF does not survive up to any finite $ T $. Of course, the BI does not  survive up to any finite $ T $ either.
   There should be a KT transition above both the SF and TSF. The $ T_{KT} $ can be estimated as $ T_{KT} \sim t \sim 3nK $
   which is clearly experimental reachable with the current cooling techniques \cite{cool1,cool2}.
%   The scaling functions across the topological phase transition driven by the Rashba SOC in a honeycomb lattice was studied in \cite{tqpt}.
   Using Eq.\ref{f33hc1} and following the procedures in \cite{tqpt}, one
   may also write down the finite temperature scaling functions for several physical quantities such as specific heats,
compressibility, Wilson ratio and thermal Hall conductivity \cite{hightc} across the $ T=0 $ SF to the $C=2 $ TSF transition near $ h_{c1} $ in Fig.1.
Following the quantum impurity problems \cite{impurity},
one may also calculate the leading corrections to the scalings
due to the leading dangerously irrelevant $ -B( q^2_x-q^2_y)^2 $ operator.

 Now we discuss the experimental detections of Fig.1 in the cold atoms.
% All the phases, bulk excitations, edge modes and phase transitions shown in Fig.1-7 are at $ T=0 $.
% But they are responsible for all the experimental measurable quantities at a finite temperature \cite{tqpt}.
 The fermionic quasi-particle spectrum can be detected by photoemission spectroscopy \cite{radio}.
 The topological phase transitions and the BCS to BEC crossovers in Fig.\ref{f33mini}
 can also be monitored by the radio-frequency dissociation spectra \cite{mitdiss,pairing}. The energy gaps of the two middle bands in Fig.\ref{f33berryhc1}b
 and Fig.\ref{f33berryhc2}b can be detected by the momentum resolved interband transitions \cite{topoexp}.
 The bulk Chern numbers can be measured by the techniques developed in \cite{newexp2}.
 The edge states can be directly imaged through Time of flight kind of measurements \cite{edgeimage}.
 A vortex can be generated by rotating the harmonic trap \cite{rotate}.
 The Majorana zero modes and the associated spin and spatial structures inside the vortex core can be imaged
 through In Situ measurements \cite{dosexp}.

%   It remains interesting to study how this order will reveal the effects of SOC.

\section{ Conclusions and Discussions }

It is constructive to compare the $ C=2 $ TSF in Fig.\ref{f33mini} with the 2d Time Reversal invariant TSF which is one copy of
2d $ p_x +i p_y $ TSF with spin up plus its Time reversal partner of a 2d $ p_x-ip_y $ with spin down\cite{zhang}.
In fact, after a unitary transformation, the surface of a 3d Topological insulator in the proximity of a S-wave superconductor
also belongs to the same class of 2d Time-reversal invariant TSF \cite{kane}.
Its two edge modes carry opposite spin and flow in opposite ( chiral ) direction. It is characterized by  the $ Z_2 $ topological invariant.
Here, the $ C=2 $ TSF has  also two copies of 2d $ p_x +i p_y $  TSF related by
the $ \mu=0 $ symmetry. However, the two edge modes are separated by the momentum $ k_y= \pi $, flow in the same ( chiral ) direction.
It is characterized by  the $ Z $ topological invariant.
As shown in the method section, the two  edge modes carry similar
spin structure near $ h_{c1} $, but opposite spin near $ h_{c2} $.

Fig.1 shows that at any value of Zeeman field, both the TSF and trivial SF are fully gapped,
there is no fermions left unpaired. This is another salient feature
due to the SOC which favors SF phase in a Zeeman field. It is  the SOC which splits the FS leading to complete pairings even in a Zeeman field.
This is in sharp contrast to S-wave pairing of spin-imbalanced fermions without SOC due to a Zeeman field where there are always
fermions left unpaired \cite{fflo,fflo1,fflo2}. The absence of FFLO state of SOC fermions with a negative interaction in a Zeeman field  reflects well the
absence of Ferromagnetic state of SOC fermions with a repulsive interaction \cite{socsdw,eta}.

 In Ref.\cite{rafhm}, we studied the same model Eq.\ref{ham} with the repulsive interaction $ U > 0 $ at a zero Zeeman field $ h=0 $.
 The positive interaction leads to spin-bond correlated magnetic phases.
 Along the extremely anisotropic line $ ( \alpha=\pi/2, \beta ) $, the ground state remains the
 $ Y-(0,\pi) $ state along the whole line and also from the weak to strong coupling.
 There is a only a crossover from the weak coupling to the strong coupling along this line.
 However, along the diagonal line $ \alpha=\beta $,
 there must be some quantum phase transitions from the $ X-(\pi,\pi) $ or $ Y-(\pi,\pi) $ spin-bond correlated magnetic
 state at weak coupling to some other spin-bond correlated magnetic states in the strong coupling \cite{rhrashba}.
 The effects of a Zeeman field in the strong repulsively interacting  limit was studied in \cite{rhh,rhtran}.
 Due to the lack of the spin $ SU(2) $ symmetry, different orientations
 of the Zeeman field lead to different phenomena \cite{rhh,rhtran}. In this paper, we only focused on the normal Zeeman field,
 it may be also interesting to study the effects of in-plane fields in Eq.\ref{ham}.

   The multi-minima structure in the ground states in Fig.\ref{f33mini} is responsible to the 1st order transition between the BI and the SF,
   also the topological first order transition from the SF and the TSF between the T and M point in Fig.\ref{f33mini}.
   It was shown in \cite{rhrashba} that the potential second order transition driven by the condensations of magnons in the Y-x state
   is pre-emptied by a first order transition between Y-x state and an In-commensurate co-planar phase in the Rotated Heisenberg model in
   the generic $ ( \alpha,\beta) $ phase diagram . These first order transitions lead to associated phase separations, meta-stable phases
   and hysteresis. In fact, the multi-minima structure also exists in the in-commensurate magnons above
   a commensurate ground state \cite{rhtran}. It is the SOC which lead to
   all these salient features in different contexts.

%All these phenomena are another salient feature due to SOC.

% For possible bosonic topological SF studied in Sec.I, the excitations may become fractionalized.
%Coulomb interactions have been addressed in \cite{qh,qhdis}.
%The P-H symmetry at the half filling $ \mu=0 $ is lost, how this P-H symmetry changes the 4 Dirac point structures
%at $ (0,0), (\pi,\pi) $ and $ (0,\pi),(\pi,0) $, especially leads to $ C=\pm 1 $ TSF and the corresponding edge mode and
%Majorana zero mode inside its vortex core will be discussed in a separate publication.
Going beyond the BCS mean field level, it maybe important to incorporate the quantum fluctuation effects.
By writing the pairing $ \Delta =\sqrt{ \Delta_0+ \delta \rho} e^{i \theta} $,
at the half filling $ \mu=0 $, we expect there exists both gapless Goldstone mode $ \theta $ and stable gapped Higgs mode $ \delta \rho $
as the collective excitation \cite{higgs} inside both the TSF and trial SF.
Near $ h_{c1} $ in Eq.\ref{f33hc1}, it is important to study the coupling between the Goldstone mode, also the Higgs mode and the gapless Dirac fermions
at $(0,\pi) $ and $ (\pi,0) $
to investigate how the gapless Goldstone mode changes the universality class of the TPT at $ h_{c1} $, also the decay rate of the Higgs mode.
Near $ h_{c2} $ in Eq.\ref{f33hc2}, following the methods developed in \cite{hightc,hightc2}, it is interesting to construct a Ginsburg Landau action
to perform a Renormalization group analysis. This action will include the bosonic sector for the superfluid order parameter
$ \Phi( \boldsymbol{r} ) $, the fermionic sector  $ \psi( \boldsymbol{r} ) $ at $ (0,0) $ and $ (\pi,\pi) $ for the topological order
and an effective $ p_x+ ip_y $ coupling between the two sectors.
%In the effective action near $ h_{c1} $ and $ h_{c2} $, the two Dirac fermions are coupled to both the Goldstone and the Higgs mode.
%The two Dirac fermion mass $ M \sim h-h_{c1} $ changes sign across the TQPT.
%It is important to  and $ T=0 $.
Between $ h_{c1} $ and $ h_{c2} $, the $ C=2 $ TSF has a fermionic gap in the bulk, but two gapless modes Eq.\ref{edge},
it maybe interesting to study how the bulk gapless Goldstone mode interacts with the two gapless edge modes.

Moving away from the half-filling, then the chemical potential $ \mu $ need to be determined self-consistently.
The $ \mu=0 $ symmetry is lost.
For general $\mu $, $\alpha, \beta $, the four $\xi _{\boldsymbol{k}
_{0}}^{2}$ in Eq.\ref{fourdirac}  could take four different values, so when tuning the Zeeman
field through $h=\sqrt{\xi _{\boldsymbol{k}_{0}}^{2}+\Delta ^{2}}$ in Eq.\ref{fourdirac}, one may drive the
system to undergo four transitions into five phases, especially  $ C=\pm 1 $ TSF. It will be discussed in a separate publication.

%It would be interesting to look at
%if there is any differences between  the Goldstone and the Higgs mode in TSF and those in trivial one.
%but I doubt there is, because the topological order can only be characterized by fermionic excitations,
%it is hard to understand so called bosonic topological order !
%Maybe bosonic Quantum Hall fluids at filling factor $ \nu=1/2m $ is an example.

{\bf Acknowledgements }

  J.Ye thank X. L. Qi for helpful discussions during his visit at KITP.  We thank W. M. Liu for encouragements
  and acknowledge AFOSR FA9550-16-1-0412 for supports.
  The work at KITP was supported by NSF PHY11-25915.

\appendix

\section{  Imposing the self-consistent equations at the Mean field calculations. }
  By introducing the pairing order parameter $ \Delta $, one can rewrite the on-site interacting term in Eq.\ref{ham} as:
\begin{equation}
H_{\Delta }=-\Delta \underset{\boldsymbol{i}}{\sum }\left( c_{\boldsymbol{i}%
\uparrow }^{\dagger }c_{\boldsymbol{i}\downarrow }^{\dagger }+c_{\boldsymbol{%
i}\downarrow }c_{\boldsymbol{i}\uparrow }\right) -\frac{ \Delta
^{2} }{U}  \label{h_delta}
\end{equation}%
  where the order parameter $\Delta $ should be determined by minimizing the free energy of the system.

 For a uniform  $ \Delta $, it is convenient to transform the operator from the real space into the momentum space:
$
c_{\boldsymbol{i}}=\frac{1}{\sqrt{L_{x}L_{y}}}\underset{\boldsymbol{k}}{\sum
}e^{i\boldsymbol{k\cdot i}}c_{\boldsymbol{k}}
$
where $\boldsymbol{k}=[\frac{\pi n_{x}}{L_{x}},\frac{\pi n_{y}}{L_{y}}]$
with $n_{x/y}=-L_{x/y},-L_{x/y}+1,\cdots L_{x/y}$, which
becomes continuous in the thermodynamic limit $L_{x/y}\longrightarrow \infty $.

Finally, we can rewrite the mean-field Hamiltonian in the Nambu representation:
\begin{widetext}
\begin{eqnarray}
H_{MF} &=&\underset{\boldsymbol{k}}{\sum }\left( \frac{1}{2}\left[
\begin{array}{cccc}
c_{\boldsymbol{k}\uparrow }^{\dagger } & c_{\boldsymbol{k}\downarrow
}^{\dagger } & c_{-\boldsymbol{k}\uparrow } & c_{-\boldsymbol{k}\downarrow }%
\end{array}%
\right] \left[
\begin{array}{cccc}
\xi _{k}-h & \Lambda _{k} & 0 & -\Delta \\
\Lambda _{k}^{\dagger } & \xi _{k}+h & \Delta & 0 \\
0 & \Delta & -\xi _{k}+h & \Lambda _{k}^{\dagger } \\
-\Delta & 0 & \Lambda _{k} & -\xi _{k}-h%
\end{array}%
\right] \left[
\begin{array}{c}
c_{\boldsymbol{k}\uparrow } \\
c_{\boldsymbol{k}\downarrow } \\
c_{-\boldsymbol{k}\uparrow }^{\dagger } \\
c_{-\boldsymbol{k}\downarrow }^{\dagger }%
\end{array}%
\right] +\xi _{k}\right) -\frac{ \Delta^{2}  }{U}
\label{nambu}
\end{eqnarray}%
\end{widetext}
   which can be diagonized by introducing two Bogoliubov quasi-particles $ \alpha _{\boldsymbol{k}\pm} $:
\begin{equation}
H_{MF}=\underset{\boldsymbol{k}}{\sum }\left[ E_{\boldsymbol{k}+}\alpha _{%
\boldsymbol{k}+}^{\dagger }\alpha _{\boldsymbol{k}+}+E_{\boldsymbol{k}%
-}\alpha _{\boldsymbol{k}-}^{\dagger }\alpha _{\boldsymbol{k}-}\right] +E_{G}
\label{hamiltonian1}
\end{equation}%
  with the quasi-particle excitation energies:
\begin{equation}
E_{\boldsymbol{k}\pm }=\sqrt{\xi _{k}^{2}+\left \vert \Lambda _{k}\right
\vert ^{2}+h^{2}+\Delta ^{2}\pm 2\sqrt{\xi _{k}^{2}\left[ \left \vert
\Lambda _{k}\right \vert ^{2}+h^{2}\right] +h^{2}\Delta ^{2}}}
\label{excitationE}
\end{equation}%
where  $ \xi _{k} =-2t\left( \cos \alpha \cos k_{x}+\cos \beta \cos k_{y}\right)
-\mu, \Lambda _{k} = 2t\left( \sin \alpha \sin k_{x}-i\sin \beta \sin k_{y}\right) $
and the ground state energy is:
\begin{equation}
E_{G}=\underset{\boldsymbol{k}}{\sum }\left[ \xi _{\boldsymbol{k}}-\frac{E_{%
\boldsymbol{k}+}+E_{\boldsymbol{k}-}}{2}\right] -\frac{ \Delta^{2}  }{U}
\label{gsenergy}
\end{equation}

At zero temperature, given the experimentally controlled parameters $ U, h $ and $ N $, one can determine
the two quantities $ \Delta, \mu $  by solving the  self-consistent equations \cite{cp}:%
\begin{eqnarray}
-\frac{\partial E_{G}}{\partial \mu } &=&N  \notag \\
\frac{\partial E_{G}}{\partial \Delta } &=&0
\label{selfconsistent}
\end{eqnarray}%

It was shown in Sec.II that the lower branch $E_{\boldsymbol{k}-}$ in Eq.\ref{excitationE} always has four
extreme points at $\left( k_{0x},k_{0y}\right) =\left( 0,0\right)
,\left( \pi ,0\right) ,\left( 0,\pi \right) ,$ and $\left( \pi ,\pi \right) $.
If there exists any gapless fermionic excitation (i.e. $E_{\boldsymbol{k}-}=0$), it must occur
at one or several of the four  $\boldsymbol{k}= \boldsymbol{k}_{0}$ where
the Eq. \ref{excitationE} simplifies to:
\begin{equation}
E_{\boldsymbol{k}_{0}-}=\left \vert \sqrt{\xi _{\boldsymbol{k}%
_{0}}^{2}+\Delta ^{2}}-h\right \vert
\label{fourdirac}
\end{equation}
 which determines the possible TPT driven the gap closing of the fermionic excitations.

 Now we focus on the half filling case.
 Using the $ \mu=0 $ symmetries of the $ E_{\boldsymbol{k}\pm } $ in Eq.\ref{excitationE},
 one can show that at the half-filling $N=L_{x}L_{y}$  ( or $\nu =\frac{1}{2}$ ),
 the chemical potential $\mu =0$ for any $\Delta $.
 So one only need to focus on the
 second self-consistent Equation in Eq.\ref{selfconsistent} to determine the $ \Delta $.
 This substantially simplifies the determination of the ground state and phase transitions shown in the Fig.\ref{f33mini}.
 It turns out that the SOC leads to highly non-trivial multi-minima landscapes in the $ ( U, h ) $ space
 shown in the Fig.\ref{f33mini}.
% The main text focused on the half filling case.
 Eq.\ref{fourdirac}'s implications on topological fermionic  transitions at $ h_{c1} $ and $ h_{c2} $ are presented in the main text.

  The bosonic transition from the BI to the SF where the fermionic excitations are always gapped are presented
  in Sec.III.

%we will describe quantitatively
%each phase and phase transition in Fig.\ref{f33mini} and Fig.\ref{f33edge}.

\section{ Most general case with  $\alpha \neq \beta  $ }

   The 2d SOC parameter $ ( \alpha, \beta) $ are experimentally tunable.
   When moving away from the isotropic limit $ \beta < \alpha $,  the $ [ C_4 \times C_4 ]_D $ symmetry
   is absent,  the $ h_{c1} $ increases to:
\begin{equation}
  h_{c1}=\sqrt{ h^2_{0} + \Delta^2 }
\end{equation}
   where $ h^2_{0} =\xi ^{2}\left( 0,\pi \right) =\xi ^{2}\left(\pi, 0 \right) =4t^{2}\left( \cos \alpha -\cos \beta \right) ^{2} $.
   It vanishes in the isotropic limit $ \alpha=\beta $ as discussed in the main text.  While $ h_{c2}= 2t ( \cos \alpha + \cos \beta ) $.

\begin{figure}[tbp]
\includegraphics[width=8cm]{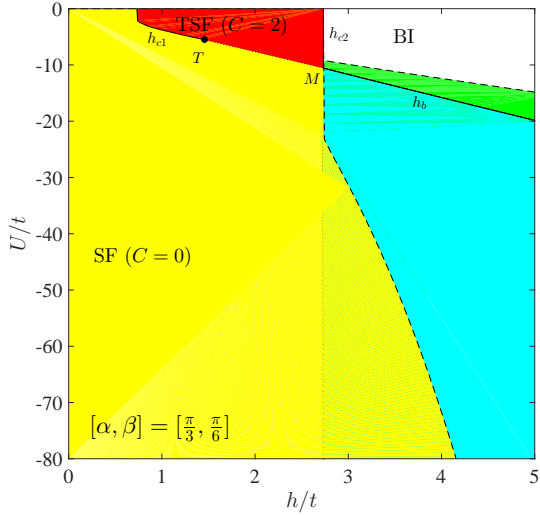}
\caption{(Color online) The global phase diagram in the parameter space of $U$ and $h$ at
$[\protect \alpha, \protect%
\beta]=[\frac{\protect \pi}{3}, \frac{\protect \pi}{6}]$. The TSF regime starts to shrink and the $ h_{c1} $ starts to approach to $ h_{c2} $.
Compare to Fig.1.}
\label{f36}
\end{figure}

   The phase diagram for $ ( \alpha=\pi/3, \beta= \pi/6 ) $ is shown in Fig.\ref{f36} where the TSF phase regime shrinks.
  Following the same procedures as those at $ \alpha=\beta $, one can derive an effective action  near $ h_{c1} $
  and near the momentum $ (0,\pi) $ or $ (\pi,0) $.
  However, due to the lack of the $ [C_4 \times C_4]_D $ symmetry at any $ \alpha \neq \beta $,
  the effective action looks more complicated than Eq.\ref{f33hc1}, but it is in the same universality class
  with a different local distribution of the Berry curvature than in Fig.2a.
  Similar statements can be made on  the effective action  near  $ h_{c2} $ and  near the momentum $ ( 0, 0)$ or $ (\pi,\pi) $.

  Note that despite the lack of $ [C_4 \times C_4]_D $ symmetry at any $ \alpha \neq \beta $, the $ \mu=0 $ symmetries remain which indicate
  the equivalence between the effective action at $ (0,\pi) $ and that at $ (\pi,0) $,
  between the effective action at $ (0,0) $ and that at $ (\pi,\pi) $ after suitable unitary transformations.
  Specific calculations showed that this is indeed the case.

\section{ The extremely anisotropic limit at  $ ( \alpha=\pi/2, \beta ) $. }

\begin{figure}[tbp]
\includegraphics[width=8cm]{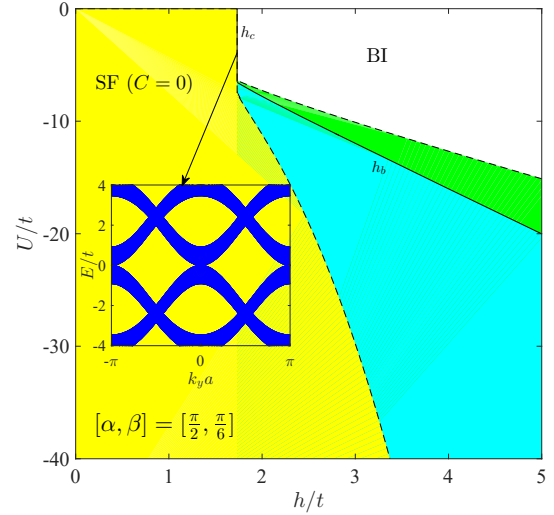}
\caption{(Color online) The global phase diagram in the parameter space of $U$ and $h$ at
$[\protect \alpha, \protect%
\beta]=[\frac{\protect \pi}{2}, \frac{\protect \pi}{6}]$.
The TSF regime shrinks to zero due to $ h_{c1}= h_{c2} =h_{c} $. Inset: the quadratic band touching at the 4 Dirac points.  }
\label{f26}
\end{figure}

It was found that in the absence of the Zeeman field, there is a spin-orbital coupled $ U(1)_{soc} $ symmetry \cite{rh} along the
anisotropic limit at  $ ( \alpha=\pi/2, \beta ) $.
The $ U(1)_{soc} $ symmetry is kept when the Zeeman field is along the $ \hat{y} $ axis \cite{rhh}.
However, it was broken when the Zeeman field is along the $ \hat{x} $ axis or the $ \hat{z} $ axis \cite{rhtran}.
Similarly, the Zeeman field along  the $ \hat{z} $ axis in Eq.M1 also breaks the $ U(1)_{soc} $.
However, as said in Sec.II, the symmetry $ \alpha \rightarrow \pi-\alpha,
k_x \rightarrow \pi-k_x $ indicates the  equivalence  between
$ (0,0) $ and $ (\pi,0) $, also between $ (0,\pi) $ and $ (\pi,\pi) $ at $ \alpha=\pi/2 $.
Then the two critical fields become the same  $ h_{c1}=h_{c2}=h_c=2t \cos \beta $.
The global phase diagram is shown in Fig.\ref{f26} where there are only two phases SF and NI, the
TSF phase is squeezed out.

The quasi-particle energy at the four points  $ (0,0), (\pi,\pi) $ and  $ ( 0,\pi), ( \pi,0) $
all  touch zero quadratically at the same time. Following the similar procedures to derive Eq.\ref{f33hc1} and Eq.\ref{f33hc2},
we reach the effective actions near the 4 points:
\begin{eqnarray}
H_{h_c} & = & \pm \left[ \delta h+\frac{\Delta ^{2}}{4t\cos \beta }-\frac{t}{%
\cos \beta }\left( q_{x}^{2}+q_{y}^{2}\right) \right] \sigma _{3}     \nonumber   \\
 & \pm &  \frac{%
\Delta }{\cos \beta }\left(  q_{x}\sigma _{1} + \sin \beta q_{y}\sigma_{2} \right )
\label{hcpi2}
\end{eqnarray}
where $ \delta h= h_c- h $ and  $  (+,-), (+,+) $ and  $  (-,+), (-,-) $  are for  $ (0,0),(\pi,\pi) $ and $ (0,\pi), (\pi,0) $ respectively.
When $ \delta h > 0 $ and $ \Delta \neq 0 $, it is in the SF phase.
When $ \delta h < 0 $ and $ \Delta =0 $, it is in the BI phase.

 In the SF side, $ \delta h > 0 $ and $ \Delta \neq 0 $, Eq.\ref{hcpi2} near any of the four points can also be cast into the form
 Eq.\ref{f33hc2}
 where $ \epsilon( \vec{k})=0 $ and $  M(\vec{k} )= M -B( q^2_x + q^2_y), M=\delta h+\frac{\Delta ^{2}}{4t\cos \beta },  B >0 $.
 The first Chern number is still given by Eq.M4:
 If $ M/B > 0 $ and $\Delta \neq 0 $ in the SF, $  C_1= \pm 1 $.  If $ M/B < 0 $ and $ \Delta=0 $ in the BI, $ C_1=0 $.
 However, the two gapped Dirac fermions at $ (0,\pi), (\pi,0) $ carry the same topological charges \cite{rafhm} $ \mu=1 $, so leading
 to the Chern number $ C_{\mu=1}=C_{(0,\pi)}+ C_{(\pi,0)}= 2 $.
 While the two gapped Dirac fermions at $ (0, 0 ), (\pi,\pi ) $ carry opposite topological charges \cite{rafhm} $ \mu=-1 $, so leading to
 to the Chern number $ C_{\mu=-1}=C_{(0,0)}+ C_{(\pi,\pi)}=- 2 $.
 So the total Chern number is $ C=2-2 =0 $, it is a trivial SF. It indicates the 4 gapped Dirac fermions can annihilate
 without going through a phase transition.

In fact, as stressed in Sec.IV and V, the topological Chern number of a given band is the integral of the Berry curvature in the
whole BZ shown in Eq.\ref{berryBZ}. Here we show that
the global topology  can be evaluated just  near a few isolated points in an effective Hamiltonian in a continuum limit.
If looking at the 4 points separately, it seems there is topological transition from a BI to a TSF with $ C=\pm 1 $.
However, the total Chern number  $ C=1+1-1-1=0 $, so globally it is still a BI to a trivial SF transition shown in Fig.\ref{f26}.

\begin{figure}[tbp]
\includegraphics[width=4.4cm]{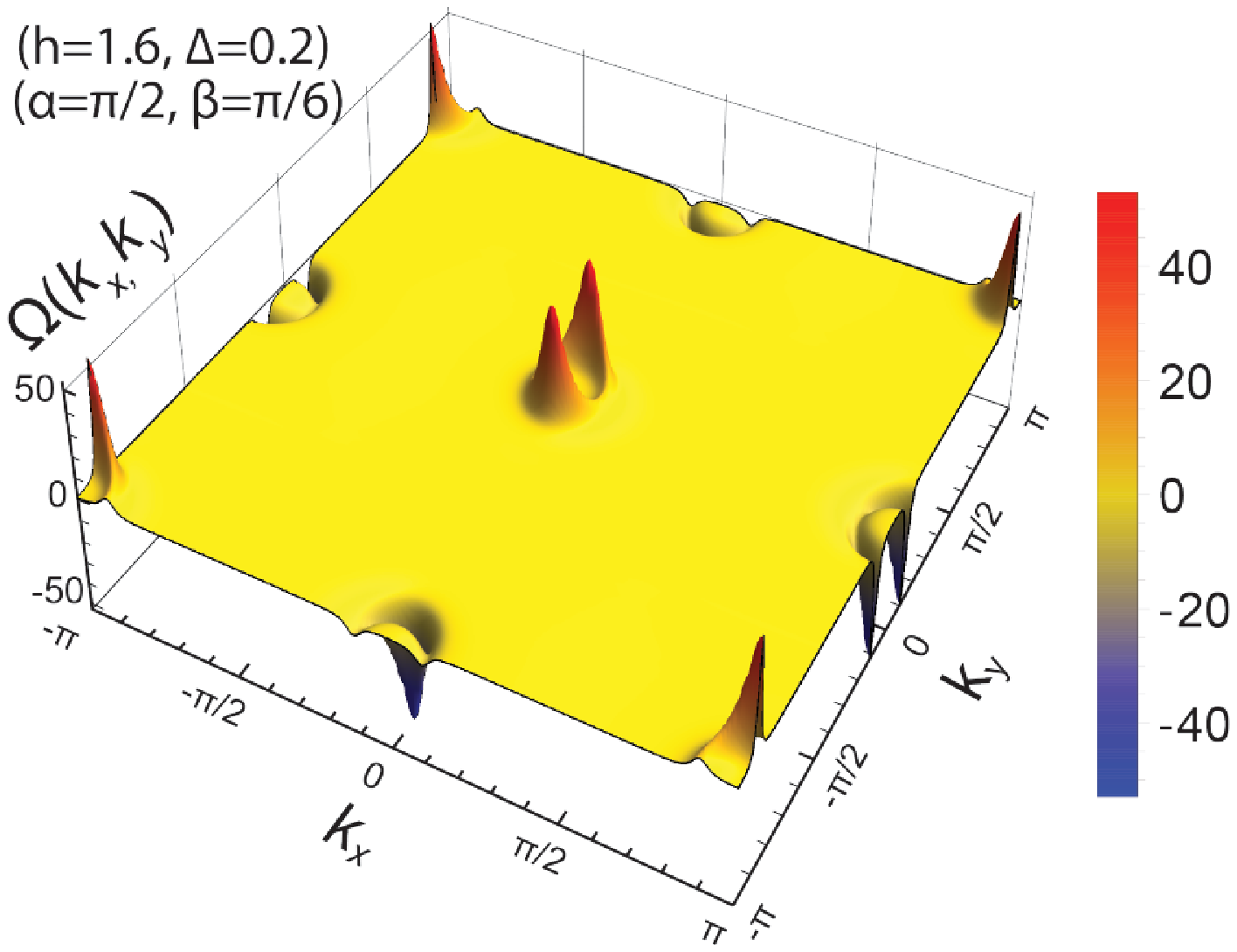}
\hspace{0.10cm}
\includegraphics[width=3.9cm]{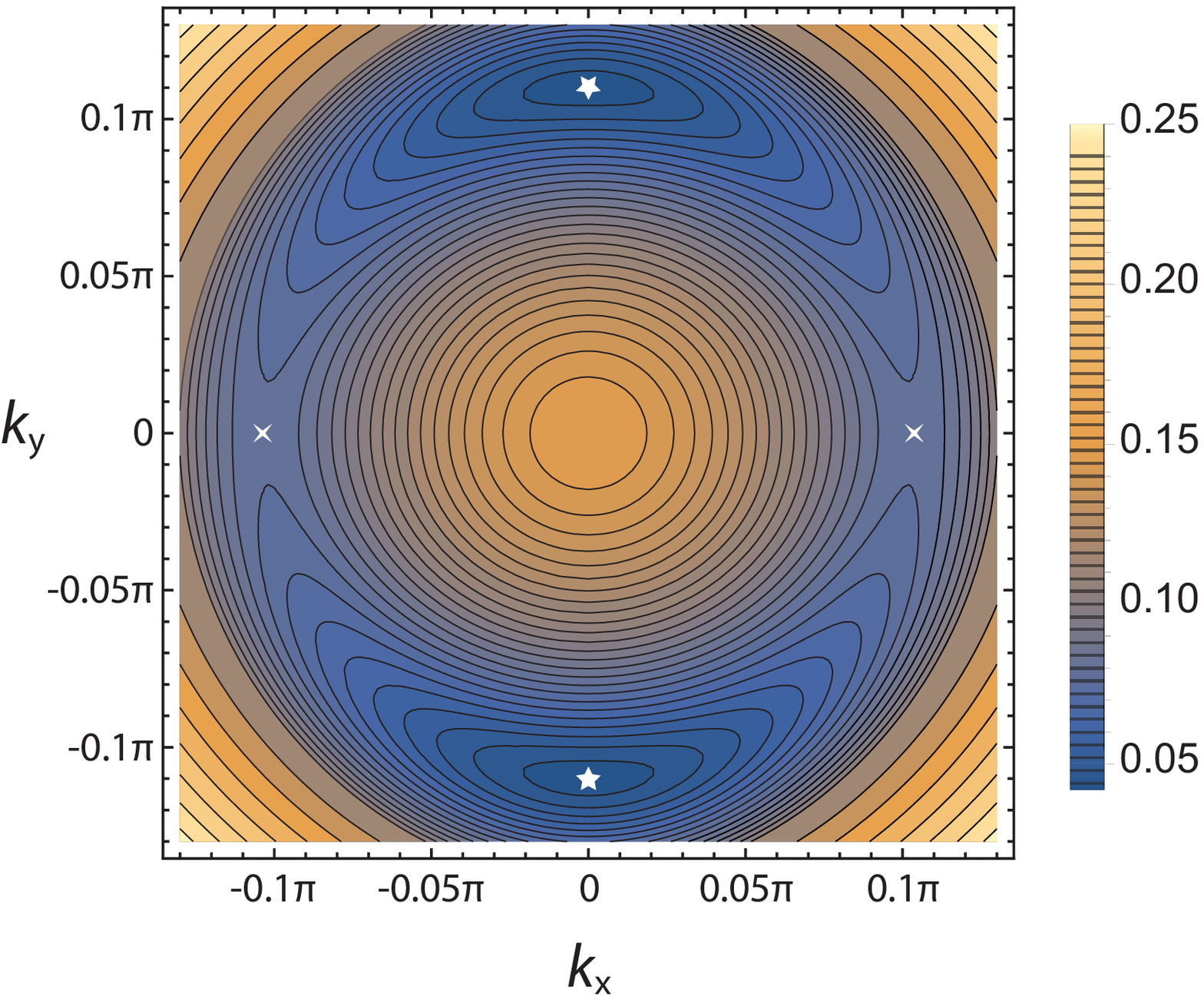}
\caption{(Color online) The Berry curvatures and energy bands  at $[\protect\alpha,
\protect\beta]=[\frac{\protect\pi}{2}, \frac{\protect\pi}{6}]$  near $ h_{c} $ with $ (h=1, \Delta=0.8 $ inside the SF.
(a) The Berry curvature has two split peaks along $ k_y $ axis  around $ (0,0), (\pi,\pi) $ with local $ C=1 $ and
$ (0,\pi), (\pi,0) $ with local $ C=-1 $.
     Despite the local non-vanishing Chern numbers, the total $ C=0 $.
(b) The energy gap contour of the two middle bands  $E_{\boldsymbol{k}-}$
 have two minima denoted by two stars along the $ k_y $ axis
    and two saddle points denoted by the two crosses along the $ k_x $ axis. This gap structure leads to that of the Berry curvature in (a). }
\label{f26berry}
\end{figure}

Using the original 4 bands theory, using three different methods outlined in Sec.IV,
we calculated the Berry Curvature of $ E_{-}( \boldsymbol{q} ) $
in the whole BZ in Fig.\ref{f26berry}a and find
they are localized around  $ (0,0),(\pi,\pi) $ and $ (0,\pi), (\pi,0) $ respectively with $ C=1,1,-1,-1 $.
The corresponding 4 energy bands are also calculated ( but not shown ).
We only draw the energy gap contour of the two middle bands  $E_{\boldsymbol{k}-}$ in Fig.\ref{f26berry}b which leads to
the non-trivial Berry curvature structure shown in Fig.\ref{f26berry}a.
All these facts can be precisely captured by the 2 bands effective theory Eq.\ref{hcpi2}.

\section{ The bulk Chern number calculations in the original 4 bands  }

  In the main text, after deriving the effective 2 band theory, we used Eq.M4 to calculate the first Chern number of a phase.
  In the following, we use 3 different methods to calculate the Berry curvature in the original 4 bands on the square lattice.
  The results are shown  in Fig.2,3 and \ref{f26berry}.
  We solve the eigenvalue problem (numerically) $	H_k|\psi_{nk}\rangle=\omega_n|\psi_{nk}\rangle $
  where the $ H_k $ is given in Eq.M11 and the eigen-energies $ \omega_1=-E_{k+},\omega_2=-E_{k-},
  \omega_3= E_{k-},\omega_4= E_{k+} $.

{\sl 1. Method 1: }
   The Berry curvature for a given band is
\begin{align}
	\Omega_n(k)=i [(\partial_x \psi_{nk}^*)(\partial_y \psi_{nk})
	-(\partial_y \psi_{nk}^*)(\partial_x \psi_{nk})]
\label{berry1}
\end{align}
    We numerically evaluate the Chern number by an integration over the whole BZ:
\begin{align}
	C_n=\frac{1}{2\pi}\int_{\rm BZ}d^2k \Omega_n (k)
\label{berryBZ}
\end{align}
Using the default numerical integration method,
we obtained $C_2=1.9999999887582889 = 2$
when $[\alpha,\beta]=[\pi/3,\pi/3]$ and $(h,\Delta)=(1,1/2)$ falling inside the TSF in
Fig.1.

 {\sl 2. Method 2: }   Eq.\ref{berry1} can also be written as
\begin{align}
	\Omega_n(k)=-\sum_{n'\neq n}
	\frac{2{\rm Im}
	\langle\psi_{nk}|(\partial_x H_k)|\psi_{n'k}\rangle
	\langle\psi_{n'k}|(\partial_y H_k)|\psi_{nk}\rangle}
	{(\omega_{n'}-\omega_n)^2}
\label{berry2}
\end{align}
%    Numerical results for Berry curvature are shown in Fig.\ref{f33berryhc1}, \ref{f33berryhc2} and \ref{f26berry}.

Setting $[\alpha,\beta]=[\pi/3,\pi/3]$.
When $(h,\Delta)=(1,1/2)$ falling in the TSF,
$C_1=-1.63415\times 10^{-12} =0$, $C_2=2.000000000001559=2$.

When  $(h,\Delta)=(1,2)$ falling in the SF,
$C_1=2.15106\times 10^{-16}=0$,
$C_2=5.99347\times 10^{-16}=0$.

When $(h,\Delta)=(4,1/2)$ falling in the BI,
$C_1=3.06829\times 10^{-16}=0$,
$C_2=-2.31586\times 10^{-16}=0$.

{\sl 2. Method 3: } This method was designed in \cite{fukui} to give exact integer Chern numbers.
One first define a U(1) link variable from the wave functions of the $n$-th band as:
\begin{align}
	U_\mu(k_l)=
	\langle \psi_{n}(k_l)|\psi_{n}(k_l+\hat{\mu}_i)\rangle/
	|\langle \psi_{n}(k_l)|\psi_{n}(k_l+\hat{\mu})\rangle|
\end{align}
where $\hat{\mu}_i$ is a vector
in the direction $i=x,y$ with the magnitude $2\pi/N_i$.
Then one define a lattice field strength as,
\begin{align}
	F_{xy}(k_l)=\ln[U_x(k_l)U_y(k_l+\mu_x)
	U_x(k_l+\mu_y)^{-1}U_y(k_l)^{-1}]
\end{align}
where the principal branch of the logarithm is with,
$-\pi<F_{xy}/i\leq\pi$.
The Chern number associated to the band $\omega_n$
is given by,
\begin{align}
	C_n=\frac{i}{2\pi}\sum_lF_{xy}(k_l)
\end{align}
It is a very efficient method. Within 10 seconds on a conventional laptop, we obtain $C_1=0$, $C_2=2$, $C_3=-2$ and $C_4=0$
when $(h,\Delta)=(1,1/2)$ falling in the TSF.

  We also used the three methods to calculate the Berry curvature using the 2 bands effective actions in Eq.\ref{f33hc1},\ref{f33hc2}
  and Eq.\ref{hcpi2} and found they reproduce those from the original four bands theory
  shown  in Fig.2, 3 and \ref{f26berry}  very precisely.

\section{ The classification of effective theories to describe 2d TPT.  }
   It is interesting to consider a generalization of the effective theories in
   Eq.\ref{f33hc1} and \ref{f33hc2}:
\begin{align}
    H_{\pm}=[\delta-(q_x^2\pm q_y^2)^n]\sigma_z+q_x\sigma_x+q_y\sigma_y
\label{qxqyn}
\end{align}
  where $ n=1,2,\cdots $ is any positive integer.

   Eq.\ref{winding} leads to the first Chern number of the lower band:
\begin{align}
    C_{\pm} =\frac{1}{4\pi}\int d^2\mathbf{q}
	\frac{\delta-(q_x^2\pm q_y^2)^n+2n(q_x^2\pm q_y^2)^n}
	   {\{[\delta-(q_x^2\pm q_y^2)^n]^2+q_x^2+q_y^2\}^{3/2}}
\label{eq:C}
\end{align}
  which can be evaluated most conveniently in the polar coordinates $  q_x=q\cos\xi,\quad q_y=q\sin\xi $.

   For the $ (q_x^2 + q_y^2)^n $ in Eq.\ref{qxqyn}, we obtain
\begin{align}
    C_{+} =\frac{1+{\rm sgn}(\delta)}{2}
\end{align}
    which leads to
\begin{eqnarray}
     C_{+} = \left \{ \begin{array}{ll}
	 1, \quad n=1,2,3,4,\cdots,~~~ \delta > 0
     \\
	 0, \quad n=1,2,3,4,\cdots,~~~ \delta < 0
    \end{array}     \right.
\label{cnp}
\end{eqnarray}

   Eq.\ref{f33hc2} realizes $ n=1 $ case in Eq.\ref{cnp}.
   It also describe the 2d TI to trivial insulator transition and the QAH to trivial insulator transition \cite{zhang}.

   For the $ (q_x^2 - q_y^2)^n $ in Eq.\ref{qxqyn}, we obtain
\begin{align}
    C_{-} =\frac{1+(-1)^n+2{\rm sgn}(\delta)}{4}
\end{align}
    If $ \delta > 0 $, it leads to
\begin{align}
	C_{-}=\begin{cases}
        1, \quad n=2,4,6,8\cdots\\
		1/2, \quad n=1,3,5,7\cdots\\	
	\end{cases}
\label{cnm1}
\end{align}
      If $ \delta < 0 $, it leads to
\begin{align}
	C_{-}=\begin{cases}
        0, \quad n=2,4,6,8\cdots\\
		-1/2, \quad n=1,3,5,7\cdots\\
	\end{cases}
\label{cnm2}
\end{align}

   Eq.\ref{f33hc2} realizes $ n=2 $ case in Eq.\ref{cnm1} and \ref{cnm2}.
   While the $ n=1 $ case in Eq.\ref{cnm1} and \ref{cnm2} describe the $ C=-1 $ QAH to $ C=1 $ QAH
   at the two Dirac fermions $ (0,\pi) $ and $ (\pi,0) $  with the same jump of the Chern number $ \Delta C= 2 $.
   In fact, when away from half filling, it was shown in \cite{un} that  the $ n=1 $ case also
   describes the TPT from $ C=-1 $ TSF to $ C=1 $  TSF  with the same jump of the Chern number $ \Delta C= 2 $.

   Note that  $ (q_x^2 - q_y^2)^n $ vanishes along the two lines $ q_{x}= \pm q_y $.
   However, due to its vanishing measure in the 2d bulk momentum space, it does not affect the total bulk Chern number $ C_{-} $.
   However, as shown in the next section,  it vanishes along the whole edges  $ q_{x}= \pm q_y $, it is not clear
   if higher order terms are needed to lead to unique edge states.


\begin{thebibliography}{99}
\bibitem{pfaffian} We expect that the Pfaffian defined from the two discrete symmetries respectively at the four momenta are identical.
It was known \cite{kane} that the Pfaffian is equal to the parity of the Chern number.
Because, the TSF in Fig.1 has Chern number $ C=2 $, so the  Pfaffian can not be used to distinguich the trvial SF with $ C=0 $ and
the TSF with $ C=2 $ in Fig.1.
\bibitem{reverse} In fact, by using the unitary transformation $ U=e^{ i \pi/2 \sigma_z } $, one can reach the same Hamiltonian
with the corresponding transformation in the two component spinor $ \phi_L $.
\bibitem{norm} Here, we are using the normalization  $ \{ \gamma_i, \gamma_j \} =\delta_{ij} $
instead of the more conventional one $ \{ \gamma_i, \gamma_j \} =2 \delta_{ij} $
which seems more natural in the BdG equation normalization.
\bibitem{eta} It was known that the finite momentum $ \eta $ pairing in the negative $ U $ Hubbard model is mapped to
a FM in XY plane in the positive $ U $ case by a particle-hole transformation.
\bibitem{cp} The chemical potential means the Fermi energy only in the normal phase where $ \Delta=0 $. However, when inside a SF,
the particle number is not conserved, it can only be used to determine the average particle number.
\end{thebibliography}
\end{document}